\documentclass[lettersize, journal, twoside]{IEEEtran}

\usepackage{etoolbox}
\usepackage{enumitem}
\usepackage[nolist,nohyperlinks]{acronym}
\usepackage{NotationMacros}
\usepackage{siunitx}
\usepackage{caption}
\usepackage{subfig}
\usepackage{adjustbox}
\usepackage{booktabs}
\usepackage{placeins}
\usepackage{orcidlink}

\usepackage{graphicx}
\usepackage{pgf, pgfplots, tikz, ifthen}
\usetikzlibrary{3d, decorations.pathreplacing, decorations.markings, shapes, arrows, arrows.meta, positioning, intersections, calc, spy, perspective, patterns, patterns.meta, plotmarks, fit, math, snakes, pgfplots.groupplots, external}
\usepgflibrary{fpu}
\usepgfplotslibrary{colorbrewer}
\pgfplotsset{compat=1.18}

\AtBeginDocument{\RenewCommandCopy\qty\SI}

\newcommand{\mcite}[1]{\mbox{\cite{#1}}}
\newcommand{\mscite}[2][]{\mbox{\cite[#1]{#2}}}

\begin{document}



\title{{Modeling nonuniform energy decay through the modal decomposition of acoustic radiance transfer (MoD-ART)}}

\author{Matteo Scerbo\,\orcidlink{0000-0001-6057-7432},
\and
Sebastian J.~Schlecht,~\IEEEmembership{Senior member,~IEEE}\,\orcidlink{0000-0001-8858-4642},
\and
Randall Ali\,\orcidlink{0000-0001-7826-1030},
\and
Lauri Savioja,~\IEEEmembership{Senior member,~IEEE}\,\orcidlink{0000-0002-8261-4596},
\and
Enzo De Sena,~\IEEEmembership{Senior member,~IEEE}\,\orcidlink{0000-0002-8007-4370}
\thanks{This work was supported by the Engineering and Physical Sciences Research Council through SCalable Room Acoustics Modelling under Grant EP/V002554/1 and through Challenges in Immersive Audio Technology under Grant EP/X032914/1. The work of Randall Ali was supported by the FWO through The Boundary Element Method as a State-Space Realization Problem under Grant G0A0424N. An earlier version of this paper was presented at the
27th International Conference on Digital Audio Effects (DAFx24).}
\thanks{Matteo Scerbo, Randall Ali, and Enzo De Sena are with the Institute of Sound Recording, University of Surrey, GU2 7XH Guildford, U.K. (e-mail: \href{mailto:m.scerbo@surrey.ac.uk}{m.scerbo}, \href{mailto:r.ali@surrey.ac.uk}{r.ali}, \href{mailto:e.desena@surrey.ac.uk}{e.desena@surrey.ac.uk}).}
\thanks{Sebastian J. Schlecht is with the Chair of Multimedia Communications and Signal Processing, University of Erlangen-Nuremberg, 91054 Erlangen, Germany (e-mail: \href{mailto:sebastian.schlecht@fau.de}{sebastian.schlecht@fau.de}).}
\thanks{Lauri Savioja is with the Department of Computer Science, Acoustics Laboratory, Aalto University, 02150 Espoo, Finland (e-mail: \href{mailto:lauri.savioja@aalto.fi}{lauri.savioja@aalto.fi}).}
}


\markboth{IEEE TRANSACTIONS ON AUDIO, SPEECH AND LANGUAGE PROCESSING, VOL. 33, 2025}%
{Scerbo \MakeLowercase{\textit{et al.}}: \@title}


\definecolor{myOrange}{RGB}{230, 159, 0} 
\definecolor{myLightBlue}{RGB}{86, 180, 233} 
\definecolor{myGreen}{RGB}{0, 158, 115} 
\definecolor{myYellow}{RGB}{240, 228, 66} 
\definecolor{myDarkBlue}{RGB}{0, 114, 178} 
\definecolor{myRed}{RGB}{213, 94, 0} 
\definecolor{myPurple}{RGB}{204, 121, 167} 

\newcommand\annotateLength[4]{
    \coordinate (shift1) at ($(#2)+(#4)$);
    \coordinate (shift2) at ($(#3)+(#4)$);
    \draw[thin, gray] (#2) -- (shift1);
    \draw[thin, gray] (#3) -- (shift2);
    \draw[latex'-latex'] (shift1) -- (shift2);
    \node[fill = white, inner sep = 0.25mm, rounded corners = 0.25mm, font = \footnotesize] (label) at ($(shift1)!0.5!(shift2)$) {#1};
}

\newcommand\HorVertJump[4]{
    \coordinate (jump) at (#3|-#4);
    \draw[-{Arc Barb[harpoon, reversed]}] (#2) -| (jump);
    \draw[{Arc Barb[harpoon, reversed, right]}#1] (jump) -- (#3);
}
\newcommand\HorVertJumpJump[5]{
    \coordinate (jump1) at (#3|-#4);
    \coordinate (jump2) at (#3|-#5);
    \draw[-{Arc Barb[harpoon, reversed]}] (#2) -| (jump1);
    \draw[{Arc Barb[harpoon, reversed, right]}-{Arc Barb[harpoon, reversed]}] (jump1) -- (jump2);
    \draw[{Arc Barb[harpoon, reversed, right]}#1] (jump2) -- (#3);
}
\newcommand\HorVertJumpJumpJump[6]{
    \coordinate (jump1) at (#3|-#4);
    \coordinate (jump2) at (#3|-#5);
    \coordinate (jump3) at (#3|-#6);
    \draw[-{Arc Barb[harpoon, reversed]}] (#2) -| (jump1);
    \draw[{Arc Barb[harpoon, reversed, right]}-{Arc Barb[harpoon, reversed]}] (jump1) -- (jump2);
    \draw[{Arc Barb[harpoon, reversed, right]}-{Arc Barb[harpoon, reversed]}] (jump2) -- (jump3);
    \draw[{Arc Barb[harpoon, reversed, right]}#1] (jump3) -- (#3);
}
\newcommand\VertJumpHor[4]{
    \coordinate (jump) at (#2|-#4);
    \draw[-{Arc Barb[harpoon, reversed]}] (#2) -- (jump);
    \draw[{Arc Barb[harpoon, reversed, right]}#1] (jump) |- (#3);
}
\newcommand\VertJumpJumpHor[5]{
    \coordinate (jump1) at (#2|-#4);
    \coordinate (jump2) at (#2|-#5);
    \draw[-{Arc Barb[harpoon, reversed]}] (#2) -- (jump1);
    \draw[{Arc Barb[harpoon, reversed, right]}-{Arc Barb[harpoon, reversed]}] (jump1) -- (jump2);
    \draw[{Arc Barb[harpoon, reversed, right]}#1] (jump2) |- (#3);
}
\newcommand\VertJumpJumpJumpHor[6]{
    \coordinate (jump1) at (#2|-#4);
    \coordinate (jump2) at (#2|-#5);
    \coordinate (jump3) at (#2|-#6);
    \draw[-{Arc Barb[harpoon, reversed]}] (#2) -- (jump1);
    \draw[{Arc Barb[harpoon, reversed, right]}-{Arc Barb[harpoon, reversed]}] (jump1) -- (jump2);
    \draw[{Arc Barb[harpoon, reversed, right]}-{Arc Barb[harpoon, reversed]}] (jump2) -- (jump3);
    \draw[{Arc Barb[harpoon, reversed, right]}#1] (jump3) |- (#3);
}

\tikzset{mymark/.style n args={2}{
        decoration={
            markings,
            mark=between positions #2 and 1 step 7mm with{%
                \tikzset{solid, every mark}\tikz@options
                \pgftransformresetnontranslations\pgfuseplotmark{#1}
            },
        },
        postaction=decorate,
        /pgfplots/legend image post style={
            mark=#1, mark options={solid}, every path/.append style={nomorepostactions}
        },
    },
}

\newcommand\drawRoom[0]{
    \coordinate (V1) at (0.0, 0.0);
    \coordinate (V2) at (0.0, 8.0);
    \coordinate (V3) at (3.99, 8.0);
    \coordinate (V4) at (4.0, 4.25);
    \coordinate (V5) at (4.01, 4.99);
    \coordinate (V6) at (8.5, 5.0);
    \coordinate (V7) at (6.0, 5.01);
    \coordinate (V8) at (6.0, 13.0);
    \coordinate (V9) at (10.0, 13.0);
    \coordinate (V10) at (10.0, 5.0);
    \coordinate (V11) at (10.0, 2.0);
    \coordinate (V12) at (4.01, 2.0);
    \coordinate (V13) at (4.0, 2.75);
    \coordinate (V14) at (3.99, 0.0);
    
    \draw[line width = 1mm] (V1) -- (V2) -- (V3) -- (V4) -- (V5) -- (V6) -- (V7) -- (V8) -- (V9) -- (V10) -- (V11) -- (V12) -- (V13) -- (V14) -- cycle;
}

\tikzset{cross/.style = {cross out, minimum size = 2*(#1-\pgflinewidth), inner sep = 0pt, outer sep = 0pt}, 
cross/.default = {0.1cm}}

\pgfplotsset{
    every non boxed y axis/.style={}
}

\begin{acronym}[MoD-ART]
\acro{VR}{virtual reality}
\acro{AR}{augmented reality}
\acro{ASR}{automatic speech recognition}
\acro{FDTD}{finite-difference time-domain}
\acro{FEM}{finite element method}
\acro{BEM}{boundary element method}
\acro{GA}{geometrical acoustics}
\acro{ISM}{image source method}
\acro{RTM}{ray tracing method}
\acro{ART}{acoustic radiance transfer}
\acro{TD-ART}{time-domain acoustic radiance transfer}
\acro{FD-ART}{frequency-domain acoustic radiance transfer}
\acro{MoD-ART}{modal decomposition of acoustic radiance transfer}
\acro{RARE}{room acoustics rendering equation}
\acro{RIR}{room impulse response}
\acro{EIR}{energy impulse response}
\acro{FIR}{finite impulse response}
\acro{IIR}{infinite impulse response}
\acro{FDN}{feedback delay network}
\acro{DFDN}{directional feedback delay network}
\acro{ARN}{acoustic rendering network}
\acro{DWN}{digital waveguide network}
\acro{DWM}{digital waveguide mesh}
\acro{WGW}{waveguide web}
\acro{SDN}{scattering delay network}
\acro{NED}{normalized echo density}
\acroplural{NED}[NEDs]{normalized echo densities}
\acro{EDC}{energy decay curve}
\acro{EDT}{early decay time}
\acro{BRDF}{bidirectional reflectance distribution function}
\acro{HRTF}{head-related transfer function}
\acro{MUSHRA}{Multiple Stimulus with Hidden Reference and Anchor}
\acro{BRAS}{Benchmark for Room Acoustical Simulation}
\acro{VBAP}{vector-based amplitude panning}
\acro{HOA}{higher-order ambisonics}
\acro{EAI}{Ehrlich-Aberth iteration}
\acro{LTI}{linear time-invariant}
\acro{ADE}{acoustic diffusion equation}
\acro{MIMO}{multiple-input, multiple-output}
\acro{SISO}{single-input, single-output}
\acro{FLOPS}{floating-point operations per second}
\end{acronym}

\maketitle

\begin{abstract}
Modeling late reverberation in real-time interactive applications is a challenging task when multiple sound sources and listeners are present in the same environment.
This is especially problematic when the environment is geometrically complex and/or features uneven energy absorption (e.g. coupled volumes), because in such cases the late reverberation is dependent on the sound sources' and listeners' positions, and therefore must be adapted to their movements in real time.
We present a novel approach to the task, named \acf{MoD-ART}, which can handle highly complex scenarios with efficiency.
The approach is based on the \acl{GA} method of \acl{ART}, from which we extract a set of energy decay modes and their positional relationships with sources and listeners.
In this paper, we describe the physical and mathematical significance of \ac{MoD-ART}, highlighting its advantages and applicability to different scenarios.
Through an analysis of the method's computational complexity, we show that it compares very favorably with ray-tracing.
We also present simulation results showing that \ac{MoD-ART} can capture multiple decay slopes and flutter echoes.
\end{abstract}




\section{Introduction}
\label{sec:introduction}

The efficient prediction of late reverberation poses a considerable challenge in the field of room acoustics modeling, as it requires high reflection orders and accurate modeling of diffuse reflections.
This endeavor is relevant in video game sound design, as well as any \ac{AR} or \ac{VR} application, where plausible auralization greatly enhances the sense of immersion~\mcite{Reverb_importance_VR} and must be performed at a sufficient speed for real-time reproduction.
Of particular interest are environments which can show different energy decay behaviors depending on the positions of sound sources and listeners, in which case the interactive adaptation of reverberation is all the more important and challenging.
We refer to these challenging cases as complex environments, on account of the complex spatial properties of their reverberation.
A classic example is that of coupled volumes, where rooms possessing different reverberation properties are connected by a series of apertures, and sources/listeners may be located in different rooms.
Note that this acoustic complexity is not always synonymous with geometric complexity: a rectangular room with very unevenly distributed energy absorption is geometrically simple, but also exhibits complex spatial reverberation.

The strict requirements of real-time late reverberation prediction make many room acoustic models unsuitable for the task.
Wave-based methods such as \ac{FDTD}~\mcite{FDTD}, \ac{FEM}~\mcite{FEM}, and \ac{BEM}~\mcite{BEM} are computationally very expensive, especially for late reverberation (i.e. broadband responses of long duration), making them difficult to use for real-time processing~\mcite{FEM_complexity}.
The same can be said of the \ac{ISM}~\mcite{ISM}, which becomes prohibitively complex for very high reflection orders, especially in nonconvex environments~\mcite{ISM_nonconvex}.
Methods like beam-tracing scale better in terms of computational complexity~\mcite{Beam_tracing_1, Beam_tracing_2}, but they still cannot model diffuse reflections, which become predominant in late reverberation~\mscite[\S4.5]{Kuttruff}.
The \ac{RTM} is better suited for late reverberation~\mcite{RTM}, modeling scattering through stochastic sampling of diffuse reflections~\mcite{Ray_scattering}~---~although this requires high numbers of rays to model complex environments reliably, which is problematic for real-time applications~\mcite{GA_overview}.
Another drawback of the \ac{RTM} is that an entirely new simulation needs to be run any time a sound source or listener moves, with very limited possibility of preserving or precomputing any non-interactive components of the simulation.
\Ac{ART} and the \ac{ADE}, on the other hand, enable the precomputation of the most important parameters for energy propagation in the environment~\mcite{RARE, Parametric_diffusion_equation}, which do not depend on the positions of sources and listeners.
This alleviates the computational load at runtime, particularly in \ac{MIMO} scenarios (i.e. more than one sound source and/or listener in the same environment), because the movement of individual sources/listeners only requires minimal updates of model parameters.
In fact, \ac{ART} can be run in real time with interactive positional updates~\mcite{FD-ART}.
Even higher real-time efficiency can be achieved by using networks of delay lines in place of convolution~\mcite{Fifty_years, Fifty_more}.
Such delay networks may be designed based on physical principles, thus implicitly performing physical modeling of room acoustics; recent works have used this approach to model specific acoustically complex environments~\mcite{atalay2022scattering, Late_inhomogeneous, alary2024designing}.


We propose an approach for late reverberation prediction, the \emph{\acf{MoD-ART}}, which offers the advantages of \ac{ART}, but with much lower computational requirements (operations and memory).
Our approach is built on the observation that \ac{ART} (in particular, \ac{TD-ART}~\mcite{TD-ART}) is a recursive \ac{LTI} system, which can be expressed by its modal decomposition~\mcite{Smith_filters}.
The outcome of the decomposition is a set of eigenvalues and eigenvectors, which respectively describe the modes' position-independent behavior and their ``coupling'' with source and listener positions~\mcite{Spectral_FTDT}.
As previously mentioned, in \ac{ART} there is a separation between the interactive elements and the parameters characterizing energy propagation in the environment.
This separation is leveraged by the modal decomposition, such that the eigenvalues are representative of the environment's energy decay modes~---~which are independent of sound sources and listeners~\mcite{Common_poles, Common_slopes_2023}.
What differentiates \ac{MoD-ART} from modal analyses such as~\mcite{Spectral_FTDT, Common_poles} is that the physical interpretations in~\mcite{Spectral_FTDT, Common_poles} deal with pressure signals and find room modes, whereas \ac{MoD-ART} deals with energy signals and finds energy decay modes.
The \ac{MoD-ART} approach also bears superficial similarities to the acoustic transfer operators proposed in~\mcite{antani2012interactive, antani2012direct}, which separate \ac{ART}'s position-independent behavior from its interactive elements.
What differentiates \ac{MoD-ART}'s operators from the ones in~\mcite{antani2012interactive, antani2012direct} is their relationship to specific acoustic properties of the analyzed environment, discussed in depth in this paper.

The groundwork for \ac{MoD-ART} was previously established in~\mcite{ART_slopes_2024}, and the approach is expanded upon here.
In this paper we present a thorough discussion of the connections between \ac{MoD-ART} parameters and physical aspects of the modeled scene; we offer insight in the multiple possible applications of \ac{MoD-ART}, and analyze its computational requirements with respect to \ac{RTM} and \ac{TD-ART}.
The efficiency of the proposed approach is attributed to two principles, both discussed in this paper:
1) a small number of energy modes is sufficient to capture all desired aspects of late reverberation,
2) interactive elements (i.e. sources, listeners) do not alter the most fundamental parameters of \ac{ART}.
We present the reasoning behind both of these points, as well as their implications for the computational requirements of \ac{MoD-ART} and for its applicability in different scenarios.

The remainder of the paper is organized as follows.
Section~\ref{sec:background} gives an overview of energy responses, \ac{TD-ART}, its interpretation as an \ac{LTI} system, and its modal decomposition.
Section~\ref{sec:proposed} discusses the physical significance of the decomposition, the rationale behind the proposed method, and some practical aspects of the approach.
Section~\ref{sec:complexity} analyzes the method's computational complexity, with a particular focus on interactive operation.
Section~\ref{sec:results} presents some simulation results which illustrate different aspects and properties of the proposed method, and validate its accuracy.
Section~\ref{sec:conclusions} concludes the paper with a summary and some final remarks.

\section{Background}
\label{sec:background}


An \ac{RIR} is a signal which entirely characterizes the propagation of sound from a given source to a given listener in a room~\mcite{Kuttruff}.
The \ac{RIR} encodes the (fixed) position and orientation of the source and listener, and the effects of their directivity, in addition to the reverberant effects of the environment.
Any anechoic sound signal (i.e. one recorded in a non-reverberant environment) may be \emph{auralized} by convolution with the \ac{RIR}, which produces the same result as if the sound had been recorded in the scenario the \ac{RIR} describes, including the specific configuration of source and listener.
The \ac{EIR}, also known as echogram, corresponds to the square of the \ac{RIR}~\mcite{Kuttruff}~---~i.e. ${\energyResponse[\sampleidx] = \abs{\pressureResponse[\sampleidx]}^2}$, where $\pressureResponse[\sampleidx]$ is the \ac{RIR} and $\energyResponse[\sampleidx]$ is the \ac{EIR}.
When it comes to late reverberation, the \ac{EIR} is sufficient for plausible auralization~\mcite{Vorlander}, even though it is impossible to achieve a perfect reconstruction of the \ac{RIR} due to the ambiguity of the sign when taking its square (i.e. the phase information is lost).
For this reason, it is commonplace to use an energy-based \ac{GA} model such as \ac{RTM} to produce an \ac{EIR}, and then to retrieve a \ac{RIR} to be used in convolution by introducing stochastic sign information~\mcite{Noise_shaping}.
In this process, known as \emph{noise-shaping}, the \ac{EIR} provides the low-varying energy behavior while the stochastic process provides sample-by-sample fluctuations.
As such, the \ac{EIR} may be evaluated at a much lower sample rate than the one used for auralization~\mscite[\S9.6]{Kuttruff}.
Noise-shaping is usually performed separately for different frequency bands, employing band-passed stochastic signals and different \acp{EIR}~\mcite{Noise_shaping}.

\begin{figure}[tb]
    \centering
    \begin{adjustbox}{max width=0.485\textwidth}
    \input{figures/volumetric_paths}
    \end{adjustbox}
    \caption{Illustration of three volumetric paths. In total, there are two volumetric paths connecting each pair of patches with an unobstructed view of each other (one path in each direction).
    The paths in this figure go from patch 1 to patch 2 ($P_{1,2}$) , from 1 to 3 ($P_{1,3}$) , and from 2 to 3 ($P_{2,3}$) .
   The figure also shows a sound source and a listener, respectively supplementing and detecting the energy contained by the volumetric paths.}
    \label{fig:volumetric paths}
\end{figure}

\subsection{Acoustic Radiance Transfer}
\label{sec:background-ART}

\Acl{ART} is a \ac{GA} model that characterizes acoustic energy as propagating over a discrete set of ``volumetric paths'' in the environment.
It is based on the assumption that energy is diffuse and uniform within each path~\mcite{RARE}, making it particularly suitable to model late reverberation (reverberant energy gradually becomes diffuse over time~\mcite{Kuttruff}).
Some examples of volumetric paths are illustrated in \figurename~\ref{fig:volumetric paths}: they connect pairs of surface patches, and they are orientation-sensitive.
Also illustrated in \figurename~\ref{fig:volumetric paths} are a sound source and a listener; their interaction with volumetric paths is explained later in this same section.

The method proposed in this paper is derived starting from \ac{TD-ART}, an implementation of \ac{ART} which produces the desired \ac{EIR} by iteratively propagating the volumetric paths' energy over (discrete) time~\mcite{TD-ART}.
The \ac{MIMO} \ac{TD-ART} model can be expressed in the $z$-domain as
\begin{subequations}
\begin{align}
    \Ssstate(z) &=
    \FeedbackMat \DelayMatArg{\feedback}(z) \Ssstate(z)
    + \Ingains(z) \Insignal(z)
    \, ,
    \label{eq:ZD-ART state}
    \\
    \Outsignal(z) &=
    \Outgains(z) \Ssstate(z)
    + \Directgains(z) \Insignal(z)
    \, ,
    \label{eq:ZD-ART output}
\end{align}
\label{eq:ZD-ART}
\end{subequations}
where ${\Insignal(z)\!\in\!\set{C}^{\numberOfSources\!\times\!1}}$ are the input energy signals (with $\numberOfSources$ sound sources), ${\Outsignal(z)\!\in\!\set{C}^{\numberOfListeners\!\times\!1}}$ are the output energy signals (with $\numberOfListeners$ listeners), ${\Ssstate(z)\!\in\!\set{C}^{\numberOfLines\!\times\!1}}$ are the energy signals traversing each of the $\numberOfLines$ volumetric paths, and the remaining parameters $\FeedbackMat$, $\DelayMatArg{\feedback}(z)$, $\Ingains(z)$, $\Outgains(z)$, and $\Directgains(z)$ are defined in the following.
The block diagram form of this system is presented in \figurename~\ref{fig:ART diagram}.

\begin{figure}[tb]
    \centering
    \begin{adjustbox}{max width=0.485\textwidth}
    \begin{tikzpicture}[auto, >=latex']
    \tikzstyle{block} = [draw, shape = rectangle, minimum height = 2em, minimum width = 3.3em, node distance = 2cm, line width = 1pt]
    
    \tikzstyle{matrix} = [draw, shape = rectangle, minimum height = 1.2cm, minimum width = 1.2cm, node distance = 2cm, line width = 1pt]
    
    \tikzstyle{sum} = [draw, shape = circle, node distance = 1.5cm, line width = 1pt, minimum width = 1.25em]
    
    \tikzstyle{branch} = [fill, shape = circle, minimum size = 0.75em, inner sep = 0pt]
    
    \node at (0, 0) (input) {\large$\Insignal(z)$};
    \node [branch, right = 0.4cm of input] (branch_in) {};
    
    \node [matrix, right = 0.5cm of branch_in] (in_gains) {\large$\Ingains(z)$};
    \node [sum, right = 0.5cm of in_gains] (sum_in) {};
    \node at (sum_in) (plus) {{\footnotesize$+$}};
    \node [branch, right = 3.7cm of sum_in] (branch_feedback) {};
    \node [matrix, right = 0.5cm of branch_feedback] (out_gains) {\large$\Outgains(z)$};
    \node [sum, right = 0.5cm of out_gains] (sum_out) {};
    \node at (sum_out) (plus) {{\footnotesize$+$}};
    
    \node [matrix, above right = 0.8cm and 0.3cm of sum_in] (A) {\large$\FeedbackMat$};
    \node [matrix, right = 0.5cm of A] (T) {\large$\DelayMatArg{\feedback}(z)$};
    
    \node [matrix, below = 0.5cm of out_gains] (bypass) {\large$\Directgains(z)$};
    \node [right = 0.5cm of sum_out] (output) {\large$\Outsignal(z)$};
    
    \node [below = -0.05cm of branch_feedback] (state) {\large$\Ssstate(z)$};
    
    \begin{scope}[line width = 1pt]
        \draw[->, double] (input) -- (in_gains);
        \draw[->, double] (branch_in) |- (bypass);
        
        \draw[->, double] (in_gains) -- (sum_in);
        
        \draw[->, double] (T) -- (A);
        \draw[->, double] (A) -| (sum_in);
        
        \draw[->, double] (branch_feedback) |- (T);
        
        \draw[->, double] (sum_in) -- (out_gains);
        
        \draw[->, double] (out_gains) -- (sum_out);
        \draw[->, double] (bypass) -| (sum_out);
        \draw[->, double] (sum_out) -- (output);
    \end{scope}

    \node at (branch_in) [branch] {};
    \node at (branch_feedback) [branch] {};
\end{tikzpicture}
    \end{adjustbox}
    \caption{Block diagram of a \ac{TD-ART} model, as expressed in~\eqref{eq:ZD-ART}.
    The filters in $\Ingains(z)$ describe how the input energy $\Insignal(z)$ is distributed among the volumetric paths, and the filters $\Outgains(z)$ describe how the volumetric paths' energy $\Ssstate(z)$ is gathered into the energy outputs $\Outsignal(z)$.
    The delays $\DelayMatArg{\feedback}(z)$ model the propagation of energy along each volumetric path, and the matrix $\FeedbackMat$ describes how energy is reflected/scattered from each path to the others.
    The filters $\Directgains(z)$ model the direct energy propagation between sources and listeners.
    }
    \label{fig:ART diagram}
\end{figure}

The matrices ${\FeedbackMat\!\in\!\set{R}^{\numberOfLines\!\times\!\numberOfLines}}$ and ${\DelayMatArg{\feedback}(z)\!\in\!\set{C}^{\numberOfLines\!\times\!\numberOfLines}}$ govern recursive energy propagation in the modeled environment.
The first, $\FeedbackMat$, describes energy reflection and diffusion among the paths; it encodes the information on the surface patches’ absorption and scattering coefficients, as well as the form factors relating pairs of patches\footnote{Also known as shape factors or view factors, these are the primary descriptors of radiance transfer~\mscite[\S2.5]{Form_factors}. They can be interpreted as the surface patches' relative ``view'' of each other, or the ``aperture'' of the volumetric propagation paths.}.
The second, $\DelayMatArg{\feedback}(z)$, is a diagonal matrix of filters applying delay and (optionally) temporal spreading~\mcite{TD-ART}, modeling propagation along the related path.
The filters forming the diagonal of $\DelayMatArg{\feedback}(z)$ may also apply air absorption.
The operation ${\FeedbackMat \DelayMatArg{\feedback}(z) \Ssstate(z)}$ in~\eqref{eq:ZD-ART state} models the energy being propagated and attenuated along each path, then reflected and mixed into the same set of paths.

The matrix ${\Ingains(z)\!\in\!\set{C}^{\numberOfLines\!\times\!\numberOfSources}}$ describes the amount of energy provided by each sound source to each propagation path, as well as the propagation delays from the source positions to the paths themselves.
In a similar fashion, ${\Outgains(z)\!\in\!\set{C}^{\numberOfListeners\!\times\!\numberOfLines}}$ describes the attenuation and delay with which energy is picked up by listeners.
Both $\Ingains(z)$ and $\Outgains(z)$ are evaluated with a single order of ray-tracing, by tracing rays from each sound source and listener position and determining the volumetric path that each ray falls into.
Lastly, ${\Directgains(z)\!\in\!\set{C}^{\numberOfListeners\!\times\!\numberOfSources}}$ models the line-of-sight component between each sound source and listener, since this would not otherwise be modeled by $\FeedbackMat$ and $\DelayMatArg{\feedback}(z)$.

Note the separation of components which depend on sound source positions (i.e. $\Ingains$, $\Directgains$), listener positions (i.e. $\Outgains$, $\Directgains$), or neither (i.e. $\FeedbackMat$, $\DelayMatArg{\feedback}$).
If a single sound source or listener changes position, only the parameters related to the moving entity need to be updated before running the filter.
This aspect comes into play in the development of our proposed method.
It is particularly advantageous in \ac{MIMO} and/or interactive scenarios, as discussed further in Section~\ref{sec:complexity-runtime}.

The \ac{TD-ART} model may also be represented in state-space form (presented in Appendix~A) or in transfer function form:
\begin{align}
    \EnergyResponseZ(z) &=
    \Outgains(z)
    \inv{\left[\eye - \FeedbackMat\DelayMatArg{\feedback}(z)\right]}
    \Ingains(z)
    + \Directgains(z)
    \, .
    \label{eq:ZD-ART transfer function}
\end{align}
This transfer function is the $z$-transform of the modeled \ac{EIR} $\EnergyResponse[\sampleidx]$.
In the following section, we see how it can be expressed in terms of its modal decomposition.



\subsection{Modal decomposition}
\label{sec:background-modal decomposition}

The transfer function of any \ac{LTI} system can be expressed in terms of poles $\pole_\modeidx$ and residues $\Residues_\modeidx$, for ${\modeidx \in 1, \dots, \numberOfModes}$, where $\numberOfModes$ is the system order.
The residues $\Residues_\modeidx$ have shape ${\numberOfListeners\!\times\!\numberOfSources}$, like the transfer function itself~---~in the \ac{SISO} case, they are scalars.
Through this decomposition, the transfer function takes the form of a sum of one-pole resonators~\mcite{Smith_filters}:
\begin{align}
    \EnergyResponseZ(z) &=
    \Directgains(z) +
    \sum_{\modeidx=1}^{\numberOfModes}
    \frac{\Residues_\modeidx}{1 - \pole_\modeidx z^{-1}}
    \, ,
    \label{eq:ZD decomposition}
\end{align}
or, expressed in the time domain,
\begin{align}
    \EnergyResponse[\sampleidx] &=
    \Directgains[\sampleidx] +
    \sum_{\modeidx=1}^{\numberOfModes}
    \Residues_\modeidx
    \pole_\modeidx^\sampleidx
    \, .
    \label{eq:TD decomposition}
\end{align}
The system poles are also known as its eigenvalues.

When analyzing systems with transfer functions of the form given in~\eqref{eq:ZD-ART transfer function}, the decomposition can be carried out as detailed in~\mbox{\mcite{FDN_modal_2019, FDN_modal_2024}}.
The system poles are the roots of the polynomial
\begin{align}
    \gcp(z) &=
    \detp{\LoopMat(z)}
    \, ,
    \label{eq:characteristic polynomial}
\end{align}
where
${\LoopMat(z) = \eye - \FeedbackMat \DelayMatArg{\feedback}(z)}$
is the transfer function of the recursive loop.
The residues are defined~\mcite{FDN_modal_2019} as
\begin{align}
    \Residues_\modeidx &=
    \frac{
        \Outgains(\pole_\modeidx)
        \adj\left(\LoopMat(\pole_\modeidx)\right)
        \Ingains(\pole_\modeidx)
    }
    {
        \tr\left(
            \adj\left(\LoopMat(\pole_\modeidx)\right)
            \LoopMat\der(\pole_\modeidx)
        \right)
    }
    \, ,
    \label{eq:residues polynomial}
\end{align}
where ${\adj(\LoopMat) = \inv{\LoopMat} \detp{\LoopMat}}$ is the adjugate of $\LoopMat$, $\tr(\LoopMat)$ is the trace of $\LoopMat$, and
\begin{align}
    \LoopMat\der(z)
    = \dv{\LoopMat(z)}{z}
    &=
    - \FeedbackMat
    \dv{\DelayMatArg{\feedback}(z)}{z}
    = - \FeedbackMat \DelayMatArg{\feedback}\der(z)
    \label{eq:matrix derivative}
\end{align}
is the element-wise derivative of $\LoopMat\der(z)$~\mcite{FDN_modal_2019}.
By definition of $\pole_\modeidx$, we have that $\LoopMat(\pole_\modeidx)$ is singular,
which in turn means its adjugate has rank one~\mcite{FDN_modal_2024}.
As such, the adjugate can be expressed as an outer product of two vectors,
\begin{align}
    \adj\left(\LoopMat(\pole_\modeidx)\right) &=
    \rightVector_\modeidx \leftVector_\modeidx\herm
    \, .
    \label{eq:adjugate as product}
\end{align}
These two vectors are the left ($\leftVector_\modeidx$) and right ($\rightVector_\modeidx$) eigenvectors related to the eigenvalue $\pole_\modeidx$, and they describe the mode's relationship with each delay line in the system~\mcite{FDN_modal_2024}.
This relationship between modes and system parameters assumes a specific meaning when the decomposed system has physical significance (like \ac{TD-ART}), and the interpretation of $\pole_\modeidx$, $\leftVector_\modeidx$, and $\rightVector_\modeidx$ is heavily dependent on the specific physical nature of the system in question~\mcite{Spectral_FTDT, Common_poles}.
The physical significance of \ac{TD-ART}'s specific modal decomposition is discussed in detail in the following section, alongside its usefulness for late reverberation modeling.

\section{Proposed method}
\label{sec:proposed}

Our proposed method, \ac{MoD-ART}, is comprised of the following steps (elaborated over the remainder of this section):
\begin{enumerate}[leftmargin=*]
    \item (Offline) Evaluate non-interactive parameters: $\FeedbackMat$, $\DelayMatArg{\feedback}(z)$.
    \item (Offline) Evaluate modal parameters: $\pole_\modeidx$, $\rightVector_\modeidx$, $\leftVector_\modeidx$.
    \item (Runtime) Evaluate interactive parameters: $\Ingains(z)$, $\Outgains(z)$.
    \item (Runtime) Evaluate residues $\Residues_\modeidx$.
    \item (Runtime) Assemble the \ac{EIR} as per~\eqref{eq:TD decomposition}.
    \item (Runtime) Retrieve the \ac{RIR}, convolve with source signals.
\end{enumerate}
The true strength of \ac{MoD-ART} lies in the fact that the modal decomposition is reduced to a very small subset of poles.
The reason for this reduction stems from the physical significance of the modal parameters $\pole_\modeidx$ and $\Residues_\modeidx$, discussed in Section~\ref{sec:proposed-significance}.
We propose two criteria for selecting the subset of poles, and present some adapted modal decomposition techniques in light of this, in Section~\ref{sec:proposed-reduction}.

\subsection{Physical significance of modal parameters}
\label{sec:proposed-significance}

When \ac{TD-ART} is viewed from the standpoint of system theory, the system's impulse response is the \ac{EIR} of the modeled environment.
It bears repeating that the \ac{EIR} is the \emph{square} of the \ac{RIR}: the modal decomposition of the \ac{EIR}, discussed here, is not to be confused with a modal decomposition of the \ac{RIR}.
When the \ac{EIR} is decomposed as in~\eqref{eq:TD decomposition}, each component
\begin{equation}
    \EnergyResponse_\modeidx[\sampleidx]
    =
    \Residues_\modeidx
    \pole_\modeidx^\sampleidx
    \label{eq:single mode response}
\end{equation}
characterizes a particular \emph{energy decay mode} of the acoustic environment.
In contrast, the individual components of the \ac{RIR}'s modal decomposition are room modes, related to the resonance frequencies of the environment (also known as eigenfrequencies).
The individual components of the \ac{EIR}'s modal decomposition have very different interpretations, which are discussed in the following.
Since the relationship between \ac{RIR} and \ac{EIR} (squared \ac{RIR}) is nonlinear, there is no straight-forward connection between the modal decomposition of one or the other.
In other words, in general, there is no one-to-one relationship between individual energy decay modes and individual room modes.

The first thing to note is that the form of~\eqref{eq:TD decomposition} is quite similar to several parametric models which describe the decay of reverberation modes~\mcite{Parametric_diffusion_equation, Common_poles, Common_slopes_2023}.
The temporal evolution of each mode~\eqref{eq:single mode response}, governed by the pole $\pole_\modeidx$, is exclusively characterized by the recursion parameters $\FeedbackMat$ and $\DelayMatArg{\feedback}(z)$.
The residues $\Residues_\modeidx$, which depend on the input-output parameters $\Ingains(z)$, $\Outgains(z)$ in addition to $\FeedbackMat$, $\DelayMatArg{\feedback}(z)$, only apply scaling (they do not control the modes' temporal evolution).
As discussed in the previous section, the recursion parameters only depend on characteristics of the acoustic environment itself: they are independent of the positions of sound sources and listeners.
By extension, the same can then be said of the energy modes' temporal evolution, which is independent of the residues.
This is also equivalent to the aforementioned parametric models, where modes' decay rates are independent of sources and listeners' positions, which in turn affect the modes' scaling.

Given their interpretation as energy quantities, all of the inputs, outputs, and state variables of \ac{ART} only take nonnegative values.
This makes \ac{ART} a \emph{positive linear system}~\mcite{Positive_LTI}, and several observations follow as consequence.
For the purpose of the present work, the most important property of positive linear systems is given by the Frobenius theorem~\mscite[Thm.~11]{Positive_LTI}.
It states that such systems always have one positive and real dominant eigenvalue $\pole_\text{F}$, also known as Frobenius eigenvalue, such that ${\abs{\pole_\modeidx} \le \abs{\pole_\text{F}}, \, \forall \modeidx}$.
The theorem also states that, among the eigenvectors associated to $\pole_\text{F}$, one is guaranteed to be positive~---~in other words, the residue associated to $\pole_\text{F}$ is always positive.
An intuitive interpretation of these statements can be gained by considering the limit
\begin{equation}
    \lim_{\sampleidx \to \infty}
    \EnergyResponse[\sampleidx]
    =
    \lim_{\sampleidx \to \infty}
    \sum_{\modeidx=1}^{\numberOfModes}
    \Residues_\modeidx
    \pole_\modeidx^\sampleidx
    =
    \Residues_\text{F}
    \pole_\text{F}^\sampleidx
    \, .
    \label{eq:Frobenius limit}
\end{equation}
As the time index tends to infinity, the pole(s) with largest magnitude dominate all others.
The \ac{EIR} is entirely nonnegative (again, because of its interpretation as an energy quantity) and therefore the dominant mode(s) must be as well.
The remainder of this discussion is divided between the physical significance of pole values, and that of residue values.

\subsubsection{Physical significance of poles}
\label{sec:proposed-significance-poles}

Since \ac{ART} is a real-valued system (i.e. its input, output, and state signals are real-valued), complex poles always appear in conjugate pairs~\mcite{Smith_filters}.
In general $z$-domain system analysis, poles lying on the real positive axis characterize non-oscillatory modes; pairs of complex poles characterize oscillatory modes; poles on the negative real axis characterize modes which oscillate at the Nyquist frequency~\mcite{Smith_filters}.
Moreover, poles on the unit circle are ``critically stable'', i.e. their modes do not increase nor decrease over time; poles with magnitude below 1 are stable (decay over time), and poles with magnitude above 1 are unstable (diverge over time).
In the context of \ac{EIR} decomposition, we can attribute physical significance to each of these behaviors.

Non-oscillatory energy decay modes are the primary concern for late reverberation.
When reverberation is analyzed by using \acp{EDC}, such modes appear as slopes (on a logarithmic scale).
In simple environments, the \ac{EDC} generally follows a single slope~\mscite[\S3.6]{Kuttruff}; the reverberation time ($\RT$) is based on this.
In the context of modal decomposition, said slope is related to the dominant pole $\pole_\text{F}$ (recall the intuition based on ${\lim_{\sampleidx \to \infty} \EnergyResponse[\sampleidx]}$).
Complex environments, such as coupled volumes or spaces with uneven energy absorption, may present multiple decay slopes~\mscite[\S5.8]{Kuttruff}.
Each slope is related to a real, positive pole (i.e. a non-oscillatory mode in the \ac{EIR} decomposition), and the slope's reverberation time is related to the pole's magnitude by the following, derived in Appendix~B:
\begin{align}
    \RT(\pole_\modeidx)
    &=
    \frac{1}{\FSe}
    \frac{\ln{10^{-6}}}{\ln{\abs{\pole_\modeidx}}}
    \, ,
    \label{eq:pole mag significance}
\end{align}
where $\FSe$ is the \ac{EIR} sample rate.
This $\RT(\pole_\modeidx)$ is also valid for oscillatory modes, where it is related to the mode's amplitude envelope~\mcite{Smith_filters}.
If the \ac{EIR} includes a constant noise floor term, it will appear in the decomposition as a real pole equal to 1: a non-oscillatory mode without decay.
A similar interpretation is used in~\mcite{Common_slopes_2023}.

Oscillatory energy decay modes may take on multiple roles in the decomposition.
In the early \ac{EIR}, they model distinct early reflections and other transient details.
In the late \ac{EIR}, oscillatory modes may appear in cases where late reverberation presents long-term oscillations such as flutter echoes.
In Section~\ref{sec:results-flutter}, we show an example where a flutter echo is characterized by a set of complex poles in the decomposition.
The relationship between the phase of a complex pole and the related mode's frequency is the same as in general $z$-domain system analysis:
\begin{align}
    \freq(\pole_\modeidx)
    &=
    \FSe
    \frac{\angle \pole_\modeidx}{2 \pi}
    \, .
    \label{eq:pole phase significance}
\end{align}
Note that poles on the positive real axis correspond to zero-frequency modes, as expected.

It is important to note that the oscillation frequency of an energy mode holds no relation to the audible frequencies in the \ac{RIR}, nor to the frequencies of room modes.
The fluctuation described by a complex energy mode takes place in the \ac{EIR}, which may describe the energy of a broadband \ac{RIR}, but it may also describe the energy in a particular frequency band of an \ac{RIR}.
For example, while considering an \ac{RIR} within the octave band centered at \qty{1}{\kilo\hertz}, one might analyze its \ac{EIR} and find an energy mode with ${\freq(\pole_\modeidx) = \qty{10}{\hertz}}$; this would indicate that the \qty{1}{\kilo\hertz} octave band's energy envelope fluctuates at a rate of \qty{10}{\hertz}.
It does \emph{not} indicate a \qty{10}{\hertz} mode in the \ac{RIR}, which would not appear in the \qty{1}{\kilo\hertz} octave band.

\begin{figure*}[t]
    \centering
    \subfloat[Dominant pole]{%
        \begin{adjustbox}{height=0.36\textwidth}%
        \begin{tikzpicture}
\begin{axis}[
colormap/RdBu,
colorbar,
colorbar style={
    y tick scale label style={xshift=0.5cm},
    ytick={-2e-2, -1e-2, 0, 1e-2, 2e-2},
    yticklabels={$-2$, $-1$, $0$, $+1$, $+2$}
},
colorbar shift/.style={xshift=-1mm},
view={0}{90},
axis equal image,
tick align=outside,
tick pos=left,
xmin = -0.1, xmax = 10.1,
ymin = -0.1, ymax = 13.1,
point meta min = -2e-2,
point meta max = 2e-2,
xlabel = {x [m]},
ylabel = {y [m]}
]

\addplot [
matrix plot*,
point meta=explicit,
mesh/check=false
] table [meta index=2] {figures/data/Residue_map_1.dat};

\drawRoom

\coordinate (Src1) at (2.1, 1.9);

\draw (Src1) node[cross=5.5pt, draw = white, line width=2pt] {};
\draw (Src1) node[cross=4pt, draw = black, line width=1pt] {};

\node[below = 6pt of Src1, fill = white, inner sep = 0.25mm, rounded corners = 0.25mm, font = \footnotesize] {Source};

\end{axis}

\end{tikzpicture}%
        \end{adjustbox}%
        \label{fig:Residue_map_1}%
    }
    \hfil
    \subfloat[Second-largest pole]{%
        \begin{adjustbox}{height=0.36\textwidth}%
        \begin{tikzpicture}
\begin{axis}[
colormap/RdBu,
colorbar,
colorbar style={
    y tick scale label style={xshift=0.5cm},
    ytick={-2e-2, -1e-2, 0, 1e-2, 2e-2},
    yticklabels={$-2$, $-1$, $0$, $+1$, $+2$}
},
colorbar shift/.style={xshift=-1mm},
view={0}{90},
axis equal image,
tick align=outside,
tick pos=left,
xmin = -0.1, xmax = 10.1,
ymin = -0.1, ymax = 13.1,
point meta min = -2e-2,
point meta max = 2e-2,
xlabel = {x [m]},
]

\addplot [
matrix plot*,
point meta=explicit,
mesh/check=false
] table [meta index=2] {figures/data/Residue_map_2.dat};

\drawRoom

\coordinate (Src1) at (2.1, 1.9);

\draw (Src1) node[cross=5.5pt, draw = white, line width=2pt] {};
\draw (Src1) node[cross=4pt, draw = black, line width=1pt] {};

\node[below = 6pt of Src1, fill = white, inner sep = 0.25mm, rounded corners = 0.25mm, font = \footnotesize] {Source};

\end{axis}

\end{tikzpicture}%
        \end{adjustbox}%
        \label{fig:Residue_map_2}%
    }
    \hfil
    \subfloat[Third-largest pole]{%
        \begin{adjustbox}{height=0.36\textwidth}%
        \begin{tikzpicture}
\begin{axis}[
colormap/RdBu,
colorbar,
colorbar style={
    y tick scale label style={xshift=0.5cm},
    ytick={-6e-2, -4e-2, -2e-2, 0, 2e-2, 4e-2, 6e-2},
    yticklabels={$-6$, $-4$, $-2$, $0$, $+2$, $+4$, $+6$}
},
colorbar shift/.style={xshift=-1mm},
view={0}{90},
axis equal image,
tick align=outside,
tick pos=left,
xmin = -0.1, xmax = 10.1,
ymin = -0.1, ymax = 13.1,
point meta min = -7e-2,
point meta max = 7e-2,
xlabel = {x [m]},
]

\addplot [
matrix plot*,
point meta=explicit,
mesh/check=false
] table [meta index=2] {figures/data/Residue_map_3.dat};

\drawRoom

\coordinate (Src1) at (2.1, 1.9);

\draw (Src1) node[cross=5.5pt, draw = white, line width=2pt] {};
\draw (Src1) node[cross=4pt, draw = black, line width=1pt] {};

\node[below = 6pt of Src1, fill = white, inner sep = 0.25mm, rounded corners = 0.25mm, font = \footnotesize] {Source};

\end{axis}

\end{tikzpicture}%
        \end{adjustbox}%
        \label{fig:Residue_map_3}%
    }
    \caption{Positional dependence of residues in a scene featuring three coupled rooms. The position of the sound source is fixed, and residues' values are plotted as a function of the listener's position.
    The three shown residues are those related to the poles with~\protect\subref{fig:Residue_map_1}~largest, \protect\subref{fig:Residue_map_2}~second-largest, and \protect\subref{fig:Residue_map_3}~third-largest magnitudes, all of which are real and positive.
    Note that the three plots' color bars cover different ranges.
    }
    \label{fig:Residue_maps}
\end{figure*}

\subsubsection{Physical significance of residues}
\label{sec:proposed-significance-residues}

Residues related to real poles are themselves real, while residues related to complex poles may be complex.
Just like for poles, residues' role in the decomposition (and their physical interpretation) can be differentiated in terms of their magnitude and phase.
Residues' magnitudes simply weight the relative modes' overall energy contribution.
If $\Residues_\modeidx$ has larger magnitude for one source-listener pair in particular, it means the $\modeidx^\text{th}$ mode has a greater influence on the \ac{EIR} for that particular pairing.
For complex poles, residues' phase has a similarly straight-forward interpretation, as it regulates the oscillatory modes' starting phase.
For real, positive poles, however, residues' phase is deserving of further discussion.
As previously mentioned, residues related to real poles are likewise real~---~but they may be negative, in which case the entire term~\eqref{eq:single mode response} is negative.
Negative modes detract energy from the decomposition, although the sum of all modes (the \ac{EIR}) is always nonnegative.
They occur in environments where the reverberation presents a fade-in period for some source-listener configurations~\mcite{Fade_in_2024}.
Examples of this are presented in Section~\ref{sec:results-transition}.

The plots in \figurename~\ref{fig:Residue_maps} show how the value of three residues changes for different positions of a listener.
The considered environment~---~which is also employed for the simulations illustrated in Section~\ref{sec:results-transition}~---~is composed of three rooms connected in sequence, with the middle room being the most reverberant.
The maps in \figurename~\ref{fig:Residue_maps} are produced by fixing the sound source position (marked by an X) in the left-most room, and evaluating the residues for listeners positioned on a grid covering the entire space.
The three shown residues are those related to the three modes with poles of largest magnitude; all three poles are real and positive, and as such, their residues always take on real values.
As predicted by the Frobenius theorem, the residue of the dominant pole (\figurename~\ref{fig:Residue_map_1}) never takes negative values.
The second and third modes' residues (\figurename~\ref{fig:Residue_map_2},~\ref{fig:Residue_map_3}), on the other hand, take on negative values in the right-most and middle rooms, respectively.
This indicates that the second mode removes energy from the \ac{EIR} when the listener is in the right-most room, and the third mode does so when the listener is in the middle room.
Finally, note how the values taken by the third mode's residue (\figurename~\ref{fig:Residue_map_3}) reach larger magnitudes than the other two.
This is because the third shown energy mode (the third-most reverberant in the decomposition) is related to the left-most room (the third-most reverberant in the scene).
The sound source is placed in the left-most room, and stimulates the related energy mode with more strength than it does the other two modes.
This manifests as the residue's larger magnitude~---~recall that the residue's magnitude controls the magnitude of the mode itself.

Combining~\eqref{eq:residues polynomial},~\eqref{eq:matrix derivative}, and~\eqref{eq:adjugate as product}, residues take the form
\begin{align}
    \Residues_\modeidx
    &=
    \frac{
        \Outgains(\pole_\modeidx)
        \rightVector_\modeidx
        \leftVector_\modeidx\herm
        \Ingains(\pole_\modeidx)
    }{
        -
        \leftVector_\modeidx\herm
        \FeedbackMat
        \DelayMatArg{\feedback}\der(\pole_\modeidx)
        \rightVector_\modeidx
    }
    \, .
\label{eq:residues}
\end{align}
This is the outer product of two vectors (and a scalar):
\begin{align}
    \Residues_\modeidx &=
    \residuesArgs{\outgain}{\modeidx}\,
    \residuesArgs{\ingain}{\modeidx}\herm\,
    \residueArgs{\feedback}{\modeidx}
    \, ,
    \label{eq:split residues total}
\end{align}
where
\begin{subequations}
\begin{align}
    \residueArgs{\feedback}{\modeidx}
    &=
    \frac{1}{
    -
    \leftVector_\modeidx\herm
    \FeedbackMat
    \DelayMatArg{\feedback}\der(\pole_\modeidx)
    \rightVector_\modeidx
    }
    \, ,
    \label{eq:split residues a}
    \\
    \residuesArgs{\ingain}{\modeidx}\herm
    &=
    \leftVector_\modeidx\herm
    \Ingains(\pole_\modeidx)
    \, ,
    \label{eq:split residues b}
    \\
    \residuesArgs{\outgain}{\modeidx}
    &=
    \Outgains(\pole_\modeidx)
    \rightVector_\modeidx
    \, .
    \label{eq:split residues c}
\end{align}
\label{eq:split residues parts}
\end{subequations}
These terms' dependency (or lack thereof) on the input-output parameters is noteworthy.
That is to say, the fact that $\residuesArgs{\ingain}{\modeidx}$ exclusively depends on input parameters ($\Ingains(z)$), and $\residuesArgs{\outgain}{\modeidx}$ exclusively depends on output parameters ($\Outgains(z)$).
In fact, there is a total separation of individual sound sources and individual listeners; each source only affects one element of $\residuesArgs{\ingain}{\modeidx}$, and each listener one element of $\residuesArgs{\outgain}{\modeidx}$.
The scalar term $\residueArgs{\feedback}{\modeidx}$ is the undriven residue~\mcite{FDN_modal_2019}, and as its name implies it is independent of the input-output parameters.
This separation has considerable repercussions on computational complexity, discussed in Section~\ref{sec:complexity}.
In terms of physical significance, the form of~\eqref{eq:split residues total} implies that each mode's positional dependence can be encapsulated into one complex scalar for each sound source, and one complex scalar for each listener.

The left and right eigenvectors $\leftVector_\modeidx$ and $\rightVector_\modeidx$ also appear separately in $\residuesArgs{\ingain}{\modeidx}$ and $\residuesArgs{\outgain}{\modeidx}$.
This reveals the eigenvectors' nature as the input-output coupling of each mode.
Consider the matrix multiplications in~\eqref{eq:split residues b}~and~\eqref{eq:split residues c}: the eigenvectors translate spatial energy distributions into modal excitations.


\subsubsection{Sample rate and fractional delays}
\label{sec:proposed-significance-samplerate}

It is worth taking a moment to discuss the values of the time delays in $\DelayMatArg{\feedback}(z)$, which model propagation times throughout the environment.
The correct modeling of reverberation, of course, relies on these propagation times being captured accurately.
Since \ac{TD-ART} is a digital filter, working in discrete time, its operation requires discrete delay lengths.
Fractional delay filters would introduce negative values, making their use in \ac{ART} problematic (the propagated signals are energy quantities, and negative values are not permitted); therefore, the delay lengths must be integer.
When \ac{TD-ART} is run conventionally, this causes a trade-off concerning the \ac{EIR} sample rate $\FSe$: low sample rates increase the rounding error in these integer delays, while high sample rates increase the computational cost of running the system (see Section~\ref{sec:complexity-runtime}).

When it comes to \ac{MoD-ART}, the same trade-off takes on a different form.
The modal decomposition approaches \ac{TD-ART} analytically, and one can analyze a \ac{TD-ART} model as if it used ideal fractional delays.
This is only possible with one of the two decomposition approaches presented in the following section.
With fractional delays, rounding error is no longer a concern, and $\FSe$ may be arbitrarily low; the trade-off is the temporal resolution of the modeled \ac{EIR}.
Unfortunately, as discussed in Section~\ref{sec:complexity-decomposition}, the decomposition approach which enables the use of ideal fractional delays is considerably more computationally expensive than the one which does not.
The effects that $\FSe$ has on poles are illustrated and discussed further in Section~\ref{sec:results-fractional}.

\subsection{Reduced mode selection}
\label{sec:proposed-reduction}

Based on the physical significance of the modal parameters, it can be deduced that a small number $\numberOfSlopes$ of modes is sufficient to model late reverberation.
Specifically, the required modes can be selected based on two criteria, presented here.
Both criteria stem from the specific focus of \ac{MoD-ART} on late reverberation, i.e., the assumption is that the earliest portion of the \ac{EIR} is modeled through other means.

The first criterion concerns poles' magnitudes, related to modes' reverberation times by~\eqref{eq:pole mag significance}.
Since modes are characterized by exponential decay, when the \ac{EIR} reaches late reverberation the faster-decaying modes are many orders of magnitude smaller than the slower-decaying ones.
As such, small poles have negligible influence on late reverberation~---~the first selection criterion disregards all poles below some given magnitude threshold.
The chosen magnitude threshold dictates the decay time threshold $\TransitionTime$ after which the reduced decomposition gives an accurate approximation of the \ac{EIR}.
If $\TransitionTime$ is low, more poles are included in the selection, and the earlier part of the \ac{EIR} is captured more accurately.
If $\TransitionTime$ is high, fewer poles are included in the selection, prioritizing the later part of the \ac{EIR}.
The implications of this choice with regards to computational complexity are discussed in Section~\ref{sec:complexity-decomposition}.

The second criterion concerns poles' phases, related to modes' frequencies by~\eqref{eq:pole phase significance}.
In most late reverberation modeling scenarios, one is only interested in modeling the non-oscillatory energy decay modes~---~i.e. only the real, positive poles are required.
In cases where more detailed late energy dynamics are desired, e.g. if flutter echoes are expected,~\eqref{eq:pole phase significance} may inform the selection of modes with different periods.

In the following, we discuss two techniques to obtain the poles; their advantages, disadvantages, and use cases.

\subsubsection{Ehrlich-Aberth iteration}
\label{sec:proposed-reduction-complex}

The first pole-finding technique we present here is \ac{EAI}, as adapted by Schlecht and Habets~\mcite{FDN_modal_2019}.
This approach was developed for systems with the same form as~\eqref{eq:ZD-ART}, although the system in~\mcite{FDN_modal_2019} was not restricted to positive values (it was not an energy model).
The pole search starts from $\numberOfModes$ initial estimates $\pole^{(0)}_\modeidx$, spaced uniformly on the unit circle.
Each estimate is then iteratively updated as
\begin{equation}
    \pole^{(\iteridx+1)}_\modeidx =
    \pole^{(\iteridx)}_\modeidx
    -
    \Delta^{(\iteridx)}_\modeidx
    \, ,
    \label{eq:correction step}
\end{equation}
where the correction term $\Delta^{(\iteridx)}_\modeidx$ is a combination of the common Newton step (as utilized in the Newton root-finding method~\mcite{Roots_of_polynomials}) and a deflation term, which penalizes pairs of estimates in close proximity~---~i.e. it prevents multiple estimates from converging to the same solution.
It bears mentioning that the Newton step requires a matrix inversion, $\inv{\LoopMat}\left(\pole^{(\iteridx)}_\modeidx\right)$; this will be relevant to the method's computational complexity, discussed in Section~\ref{sec:complexity-decomposition}.
We refer to~\mcite{FDN_modal_2019} for a full discussion of these terms and the overall process.

The iteration as carried out in~\mcite{FDN_modal_2019} aims to achieve a full decomposition (i.e. finding all $\numberOfModes$ poles), and as such, it terminates on convergence of all estimates.
Convergence is assessed based on two criteria: if the latest step $\abs{\pole^{(\iteridx)}_\modeidx - \pole^{(\iteridx-1)}_\modeidx}$ is smaller than some threshold,
and/or if the condition number $\kappa(\LoopMat(\pole^{(\iteridx)}_\modeidx))$ is larger than some threshold,
the estimate $\pole^{(\iteridx)}_\modeidx$ is no longer updated.
For our purposes, as previously discussed, a full decomposition is not desired~---~only poles with significant magnitude are sought.
As such, we introduce a third stopping criterion: iteration is halted when estimates fall below a magnitude threshold.
From~\eqref{eq:pole mag significance},
\begin{equation}
    \abs{\pole^{(\iteridx)}_\modeidx} <
    10^\frac{-6}{\TransitionTime \FSe}
    \, ,
    \label{eq:EAI T60 thresh}
\end{equation}
where $\TransitionTime$ is the desired time threshold, as discussed above.

An interesting aspect of the \ac{EAI} approach is that it finds the roots of the polynomial~\eqref{eq:characteristic polynomial} even if the polynomial describes a filter that cannot be implemented in practice, such as one involving noncausal or ideally fractional delays.
Performing the decomposition with ideally fractional delays is a feature with multiple advantages, discussed in further detail in Section~\ref{sec:results-fractional}.

\subsubsection{Arnoldi iteration}
\label{sec:proposed-reduction-arnoldi}

The system poles may also be found as eigenvalues of the state transition matrix, detailed in Appendix~A.
Unlike \ac{EAI}, this approach precludes the use of ideal fractional delays in the analysis, because constructing the state transition matrix is only possible if the propagation delays are integer.
Furthermore, the eigenvalue decomposition cannot be exclusively targeted to the real axis, making this approach less fitting in cases where oscillatory modes are irrelevant.
Nevertheless, this approach is worth mentioning here, because of its higher efficiency.
Thanks to the very high sparsity of the state transition matrix and to the fact that only a few eigenvalues are desired, the decomposition can be performed using the Arnoldi iteration method~\mcite{Arnoldi}.
The comparative computational complexity of this approach with respect to \ac{EAI} is discussed further in the following section.



\section{Computational complexity}
\label{sec:complexity}

The computational costs of \ac{MoD-ART} can be differentiated between precomputation (offline) and interactive (runtime) components, as listed at the start of Section~\ref{sec:proposed}.
Since \ac{MoD-ART} is intended for real-time interactive applications, where runtime efficiency is paramount, this section prioritizes the discussion of interactive elements.
Some aspects of the precomputation are also briefly discussed, but a more extensive analysis is left for future works.
Moreover, we do not account for costs which are common between our method and other modeling approaches, e.g. the cost of noise-shaping, the cost of convolution, and the process's repeated application for different frequency bands.

\subsection{Evaluation of $A$}
\label{sec:complexity-matrix}

The \ac{ART} matrix $\FeedbackMat$ can be found by analytical means in some situations~\mcite{configuration_factors}, but it is usually evaluated through ray-tracing~\mcite{RARE}.
The leading factor in the complexity of \ac{ART}, and by extension of \ac{MoD-ART}, is the number of volumetric paths $\numberOfLines$.
This is related to the number of discrete surface patches and to the geometrical complexity of the environment.
Let us say, for example, that each of the $\numberOfPolys$ polygons making up the environment mesh is taken to be a discrete surface patch, and that each one (on average) has visibility on $\visibilityRatio \numberOfPolys$ other polygons, with ${0\!<\!\visibilityRatio\!<\!1}$.
Then, the number of volumetric paths is ${\numberOfLines = \numberOfPolys (\visibilityRatio \numberOfPolys)}$, because paths only exist between mutually visible surface patches.
The value $\visibilityRatio$ is low when the environment is highly geometrically complex, and can only equal 1 if each surface patch has visibility on all surface patches~---~meaning the enclosure is strictly convex, and surface patches have self-visibility (impossible when they are planar polygons).
Consider, for example, an environment with 10 rooms of 100 polygons each (${\numberOfPolys = 1000}$), such that each polygon (on average) has visibility on roughly 100 polygons in the same room and 20 polygons from adjacent rooms.
Then, we have a visibility ratio of ${\visibilityRatio = 120/1000 = 0.12}$, and ${\numberOfLines = 1000 \cdot 120 = 1.2\!\cdot\!10^5}$ volumetric paths.

In many cases, the environment mesh features a high number of small polygons.
For the purpose of \ac{ART}, it is desirable to have fewer, larger patches, to reduce the runtime complexity at the cost of a slightly coarser approximation.
Reducing the polygon count of the environment mesh is possible to some extent~\mcite{Geometry_reduction}, but some important features of the space may be lost in doing so.
Another option is to define discrete surface patches which span several adjacent polygons, possibly allowing non-planar surface patches.
This approach has the potential to greatly decrease the size $\numberOfLines$ of the \ac{ART} matrix, and therefore the computational complexity of the modal decomposition and of the runtime elements.
In the following, to avoid confusion, we maintain the assumption of each patch being a single polygon, and use $\numberOfPolys$ to denote the number of discrete surface patches.

Another crucial factor in the complexity of \ac{ART} is the number of nonzero elements of $\FeedbackMat$, which is a very sparse matrix.
The sparsity of $\FeedbackMat$ is due to the fact that it only describes valid reflections.
Take for example a path connecting two surface patches $a$ and $b$: the energy traversing this path will be reflected by $b$, therefore it may only be redirected to paths that start from $b$.
Let us say, as before, that $\visibilityRatio \numberOfPolys$ is the number of paths starting from any given surface patch; the number of nonzero elements of $\FeedbackMat$ is then ${\numberOfLines (\visibilityRatio \numberOfPolys)}$, i.e. ${\numberOfPolys (\visibilityRatio \numberOfPolys)^2}$.
This corresponds to just a fraction ${1/\numberOfPolys}$ of the total number of elements (including zeros) of $\FeedbackMat$, which is ${\numberOfLines^2 = \numberOfPolys^2 (\visibilityRatio \numberOfPolys)^2}$.

\subsection{Modal decomposition}
\label{sec:complexity-decomposition}

The bulk of decomposition costs relates to finding the poles, rather than the eigenvectors, which are found from the adjugate\footnote{Since the adjugate is the outer product of the eigenvectors (see~\eqref{eq:adjugate as product}), they are easily derived from the first row and first column of the adjugate.} once the poles are known.
Here, we briefly compare the two pole-finding techniques presented in Section~\ref{sec:proposed-reduction}, in terms of computational complexity.

\subsubsection{EAI approach}

The crux of \acs{EAI}'s complexity lies in the matrix inversion required by the Newton step, for each estimate, at each iteration.
Unlike other parts of the iterative process, matrix inversion cannot benefit from the sparse nature of the matrix $\LoopMat(z)$.
In addition to this, the process must start with all $\numberOfModes$ estimates (i.e. as many as the system order) in order for the deflation process to work as intended.
As such, the complexity of a single iteration (updating all estimates) is
\begin{equation}
    O\left(
    \numberOfModes^2
    +
    \numberOfModes
    \numberOfLines^3
    \right)
    \, .
    \label{eq:EAI complexity}
\end{equation}
Further details are given in~\mcite{FDN_modal_2019}.
The system order $\numberOfModes$ is always larger than the number of paths $\numberOfLines$, but (in \ac{TD-ART} models) the term $\numberOfModes \numberOfLines^3$ is usually much larger than $\numberOfModes^2$.


\subsubsection{Arnoldi approach}

The Arnoldi iteration method for eigenvalue decomposition has two main advantages, in terms of efficiency: it allows searching for a small subset of eigenvalues, and it exploits matrix sparsity~\mcite{Arnoldi}.
The subset of eigenvalues can be selected through different priority criteria, such as larger or smaller amplitude~\mcite{ARPACK}; as with the \ac{EAI} approach, the search can be halted when the desired magnitude threshold is reached.
The leading term in the computational complexity of Arnoldi iteration is related to a matrix-vector multiplication between the matrix to be decomposed (in this case, the state transition matrix) and a candidate eigenvector~\mcite{ARPACK}.
The cost of such a multiplication is proportional to the number of nonzero elements in the matrix.
For a dense matrix, this would be the square of its size.
The state transition matrix, however, is exceedingly sparse: despite having a size of ${\numberOfModes\!\times\!\numberOfModes}$, its nonzero elements are ${(\numberOfModes - \numberOfLines)}$ plus the nonzero elements of $\FeedbackMat$ (see Appendix~A).
As previously stated, $\FeedbackMat$ has ${\numberOfLines (\visibilityRatio \numberOfPolys)}$ nonzero elements, and as such, the state transition matrix has $\numberOfModes - \numberOfLines + \numberOfLines (\visibilityRatio \numberOfPolys)$ nonzero elements.
In conclusion, if $\numberOfSlopes$ poles are desired, the complexity of the Arnoldi approach is
\begin{equation}
    O\left(
    \numberOfSlopes
    \left(
    \numberOfModes
    + \numberOfLines
    (\visibilityRatio \numberOfPolys - 1)
    \right)
    \right)
    \, .
    \label{eq:Arnoldi complexity}
\end{equation}

\subsection{Interactive elements}
\label{sec:complexity-runtime}

In the following, we assume that all non-interactive parameters (i.e. poles $\pole_\modeidx$, eigenvectors $\leftVector_\modeidx$, $\rightVector_\modeidx$, and undriven residues $\residueArgs{\feedback}{\modeidx}$) have been precomputed.
What remains to be done is evaluating the input-output parameters $\Ingains(z)$, $\Outgains(z)$ and the residues $\Residues_\modeidx$.
This process may be repeated at fixed intervals, or only upon movement of the sources/listeners.

\subsubsection{Ray-tracing}

As introduced in Section~\ref{sec:background-ART}, the input-output parameters of \ac{ART} are evaluated through a single order of ray-tracing from each source, and from each listener.
For the sake of comparison against a full \ac{RTM} approach, let us discuss the complexity of ray-tracing for an arbitrary reflection order $\numberOfReflections$.
In a ``naive'' implementation of the method, each of the $\numberOfListeners$ listeners casts $\numberOfRays$ rays, and each ray (at each reflection) checks for intersections with all objects: $\numberOfPolys$ polygons, and $\numberOfSources$ sources (spherical interceptors).
In short, the complexity is
\begin{equation}
    O\left(
    \numberOfListeners
    \numberOfReflections
    \numberOfRays
    (\numberOfPolys + \numberOfSources)
    \right)
    \, .
    \label{eq:naive RTM}
\end{equation}
In optimized \ac{RTM} implementations such as KD-Tree traversal~\mcite{KD-tree_RTM}, polygons are organized into a structure that greatly expedites the intersection process, and the complexity becomes
\begin{equation}
    O\left(
    \numberOfListeners
    \numberOfReflections
    \numberOfRays
    (\log{(\numberOfPolys)} + \numberOfSources)
    \right)
    \, .
    \label{eq:Tree RTM}
\end{equation}

\begin{figure*}[t]
    \centering
\begin{tikzpicture} 
\begin{axis}[%
hide axis,
xmin=10,
xmax=50,
ymin=0,
ymax=0.4,
legend columns = 5,
legend style={draw=white!15!black,legend cell align=left}
]
\addlegendimage{myOrange, very thick, mark=star}
\addlegendentry{RTM, $\numberOfReflections = 10$}
\addlegendimage{myLightBlue, very thick, mark=triangle}
\addlegendentry{RTM, $\numberOfReflections = 100$}
\addlegendimage{myRed, very thick, mark=o}
\addlegendentry{TD-ART}
\addlegendimage{myGreen, very thick, mark=square}
\addlegendentry{TD-ART, static sources}
\addlegendimage{myPurple, very thick, mark=+}
\addlegendentry{MoD-ART}
\end{axis}

\end{tikzpicture}%
    \vfil
    \subfloat[]{%
        \begin{adjustbox}{height=0.35\textwidth}%
        \begin{tikzpicture}[]
\pgfmathsetmacro{\Poly}{140}
\pgfmathsetmacro{\Envc}{0.4}
\pgfmathsetmacro{\Rays}{100000}
\pgfmathsetmacro{\Samp}{2000}
\pgfmathsetmacro{\Pole}{10}
\begin{loglogaxis}
[
width = 10.5cm,
height = 7.5cm,
legend cell align = {left},
legend columns = 2,
legend style = {
    at = {(0.02, 0.98)},
    anchor = north west,
    fill opacity = 0.3,
    draw opacity = 1,
    text opacity = 1,
    draw = lightgray
},
tick align = outside,
tick pos = left,
xlabel = {Number of sources and listeners},
ylabel = {Computational complexity},
domain = 1:100,
xmin = 1, xmax = 100,
ymin = 1e6, ymax = 1e11,
xtick = {1, 10, 100},
xticklabels = {1, 10, 100}
]

\addplot
[
semithick,
forget plot,
mymark={o}{3mm},
myRed
]
{(2*x*\Rays*ln(\Poly)) + \Samp*(\Poly*\Poly*\Envc)*(2*x*\Envc+\Poly*\Envc)};

\addplot
[
semithick,
forget plot,
mymark={square}{4mm},
myGreen
]
{(x*\Rays*ln(\Poly)) + \Samp*(\Poly*\Poly*\Envc)*x*\Envc};

\addplot
[
semithick,
forget plot,
mymark={+}{5mm},
myPurple
]
{(2*x*\Rays*ln(\Poly)) + \Pole*(\Poly*\Poly*\Envc)*2*x*\Envc};


\addplot
[
semithick,
forget plot,
mymark={star}{1mm},
myOrange
]
{x*10*\Rays*(ln(\Poly)+x)};

\addplot
[
semithick,
forget plot,
mymark={triangle}{2mm},
myLightBlue
]
{x*100*\Rays*(ln(\Poly)+x)};


\end{loglogaxis}

\end{tikzpicture}%
        \end{adjustbox}%
        \label{fig:Complexity_triple}%
    }
    \subfloat[]{%
        \begin{adjustbox}{height=0.35\textwidth}%
        \begin{tikzpicture}[]
\pgfmathsetmacro{\Agnt}{10}
\pgfmathsetmacro{\Rays}{100000}
\pgfmathsetmacro{\Samp}{2000}
\pgfmathsetmacro{\Pole}{10}
\begin{loglogaxis}
[
width = 10.5cm,
height = 7.5cm,
legend cell align = {left},
legend style = {
    at = {(0.02, 0.98)},
    anchor = north west,
    fill opacity = 0.3,
    draw opacity = 1,
    text opacity = 1,
    draw = lightgray
},
tick align = outside,
tick pos = left,
xlabel = {Number of polygons},
domain = 10:10000,
xmin = 10, xmax = 10000,
ymin = 1e6, ymax = 1e11
]

\pgfmathsetmacro{\Envc}{0.25}

\addplot
[
semithick,
forget plot,
mymark={o}{5mm},
myRed
]
{(2*\Agnt*\Rays*ln(x)) + \Samp*(x*x*\Envc)*(2*\Agnt*\Envc+x*\Envc)};

\addplot
[
semithick,
forget plot,
mymark={square}{6mm},
myGreen
]
{(\Agnt*\Rays*ln(x)) + \Samp*(x*x*\Envc)*\Agnt*\Envc};

\addplot
[
semithick,
forget plot,
mymark={+}{7mm},
myPurple
]
{(2*\Agnt*\Rays*ln(x)) + \Pole*(x*x*\Envc)*2*\Agnt*\Envc};


\pgfmathsetmacro{\Envc}{0.5}

\addplot
[
semithick,
densely dashed,
forget plot,
mymark={o}{4mm},
myRed
]
{(2*\Agnt*\Rays*ln(x)) + \Samp*(x*x*\Envc)*(2*\Agnt*\Envc+x*\Envc)};

\addplot
[
semithick,
densely dashed,
forget plot,
mymark={square}{5mm},
myGreen
]
{(\Agnt*\Rays*ln(x)) + \Samp*(x*x*\Envc)*\Agnt*\Envc};

\addplot
[
semithick,
densely dashed,
forget plot,
mymark={+}{6mm},
myPurple
]
{(2*\Agnt*\Rays*ln(x)) + \Pole*(x*x*\Envc)*2*\Agnt*\Envc};


\pgfmathsetmacro{\Envc}{1.0}

\addplot
[
semithick,
densely dotted,
forget plot,
mymark={o}{4mm},
myRed
]
{(2*\Agnt*\Rays*ln(x)) + \Samp*(x*x*\Envc)*(2*\Agnt*\Envc+x*\Envc)};

\addplot
[
semithick,
densely dotted,
forget plot,
mymark={square}{3mm},
myGreen
]
{(\Agnt*\Rays*ln(x)) + \Samp*(x*x*\Envc)*\Agnt*\Envc};

\addplot
[
semithick,
densely dotted,
forget plot,
mymark={+}{5mm},
myPurple
]
{(2*\Agnt*\Rays*ln(x)) + \Pole*(x*x*\Envc)*2*\Agnt*\Envc};


\addplot
[
semithick,
forget plot,
mymark={star}{1mm},
myOrange
]
{\Agnt*10*\Rays*(ln(x)+\Agnt)};

\addplot
[
semithick,
forget plot,
mymark={triangle}{2mm},
myLightBlue
]
{\Agnt*100*\Rays*(ln(x)+\Agnt)};

\addplot[black, very thick, densely dotted]
table {%
1 1
2 1
};
\addlegendentry{{$\visibilityRatio = 1$}}
\addplot[black, very thick, densely dashed]
table {%
1 1
2 1
};
\addlegendentry{{$\visibilityRatio = 0.5$}}
\addplot[black, very thick]
table {%
1 1
2 1
};
\addlegendentry{{$\visibilityRatio = 0.25$}}

\end{loglogaxis}

\end{tikzpicture}%
        \end{adjustbox}%
        \label{fig:Complexity_visibility}%
    }
    \caption{Computational complexity comparison of \ac{RTM}, \ac{TD-ART}, and \ac{MoD-ART}, as discussed in Section~\ref{sec:complexity-conclusions}.
    All methods consider interactively updated source and listener positions, except for ``\ac{TD-ART}, static sources'' where only listeners are updated.
    Complexity is plotted in~\protect\subref{fig:Complexity_triple} as a function of sources and listeners, considering the three-room case shown in \figurename~\ref{fig:environment} in terms of numbers of polygons and visibility (${\numberOfPolys = 140}$, ${\visibilityRatio \approx 0.4}$).
    In~\protect\subref{fig:Complexity_visibility}, complexity is plotted as a function of the number of polygons, with ${\numberOfSources = \numberOfListeners = 10}$ sources and listeners, in three hypothetical environments with different visibility factors: ${\visibilityRatio = 0.25}$ (solid lines, low visibility), ${\visibilityRatio = 0.5}$ (dashed lines, moderate visibility), and ${\visibilityRatio = 1}$ (dotted lines, strictly convex enclosure).
    }
    \label{fig:Complexity}
\end{figure*}

Note that, if any of the sources move, the tracing process needs to be re-run from all listener positions.
If all sources are static and some listeners move, the process needs to be re-run only for moving listeners.
On the contrary, for the purpose of computing $\Ingains(z)$ and $\Outgains(z)$ for \ac{ART}, all sources and listeners are independent: if a single source or listener moves, only the parameters related to that individual need to be updated.
The tracing process then only needs to be performed from the positions of recently moved sources/listeners, with the complexity of a single order of ray-tracing from each position:
\begin{equation}
    O\left(
    \moved{\numberOfSources}
    \numberOfRays
    \log{(\numberOfPolys)}
    +
    \moved{\numberOfListeners}
    \numberOfRays
    \log{(\numberOfPolys)}
    \right)
    \, ,
    \label{eq:ART tracing cost}
\end{equation}
where $\moved{\numberOfSources}$ and $\moved{\numberOfListeners}$ refer to the number of sources and listeners which have moved since the last update, respectively.
This is equivalent to performing~\eqref{eq:Tree RTM} from each new position, with ${\numberOfReflections=1}$, and without considering intersections with sources.

\subsubsection{Running TD-ART}

The standard operation of \ac{TD-ART}, once its parameters have been prepared, is equivalent to running the digital filter described by~\eqref{eq:ZD-ART}.
The main cost is that of the matrix-vector multiplication ${\FeedbackMat \Ssstate}$, which has complexity $O\left(\numberOfLines \visibilityRatio \numberOfPolys\right)$~---~the number of nonzero elements in $\FeedbackMat$, as previously discussed.
In addition to this, there are the multiplications with $\Ingains$ and $\Outgains$, which are also sparse.
Indeed, like $\FeedbackMat$, the input-output operators $\Ingains$ and $\Outgains$ are also affected by the visibility parameter $\visibilityRatio$.
As previously stated, only ${\visibilityRatio \numberOfPolys}$ polygons are visible (on average) from any given position in the environment.
If a sound source has visibility on ${\visibilityRatio \numberOfPolys}$ surface patches, then it can only contribute energy to the ${\visibilityRatio \numberOfLines}$ volumetric paths which start from those visible patches; equivalently, a listener can only detect energy from ${\visibilityRatio \numberOfLines}$ volumetric paths.
As such, the number of nonzero elements in $\Ingains$ is ${\numberOfSources \visibilityRatio \numberOfLines}$, and those in $\Outgains$ are ${\numberOfListeners \visibilityRatio \numberOfLines}$.
In conclusion, the multiplications with $\Ingains$ and $\Outgains$ have complexity ${O\left(\numberOfSources\visibilityRatio\numberOfLines + \numberOfListeners\visibilityRatio\numberOfLines\right)}$.
All of these operations need to be performed for every sample of the \ac{EIR}, making the overall complexity
\begin{equation}
    O\left(
    \numberOfSamples
    \numberOfLines
    \visibilityRatio
    \left(
    \numberOfSources
    +
    \numberOfListeners
    +
    \numberOfPolys
    \right)
    \right)
    \, ,
    \label{eq:ART running cost}
\end{equation}
where $\numberOfSamples$ is the length of the desired \ac{EIR}.

Since \ac{TD-ART} is a time-iterative process, all of the aforementioned matrix multiplications need to be performed for each time sample of the \ac{EIR}.
The time delays in $\DelayMatArg{\feedback}(z)$ and in each filter do not constitute a computational requirement in terms of operations, but in terms of memory space; there are two ways to handle them.
The most common approach~\mcite{TD-ART} is to instantiate a matrix of size ${\numberOfLines\!\times\!\numberOfSamples}$, and use it to store the energy $\Ssstate(z)$ over time.
The advantage of this approach is that, if listeners move but sound sources are static, only the multiplications with $\Outgains$ need to be performed interactively~\mcite{FD-ART}.
An alternative approach is to implement \ac{TD-ART} as a filter similar to a \acf{FDN}~\mcite{FDN_Jot}.
This way, the iteration needs to be run interactively even if sources are static, but the memory requirements are reduced to the lengths of the propagation paths~\mcite{Schroeder}.

\subsubsection{Running MoD-ART}

The \ac{MoD-ART} approach requires the computation of the input-output weights $\Ingains(z)$ and $\Outgains(z)$ with complexity~\eqref{eq:ART tracing cost}, but it avoids the runtime cost~\eqref{eq:ART running cost}.
Once all \ac{ART} parameters have been prepared (for $\numberOfSlopes$ selected poles), the residue components $\residuesArgs{\ingain}{\modeidx}$ and $\residuesArgs{\outgain}{\modeidx}$ are computed with the dot products in~\eqref{eq:split residues parts}, with complexity
\begin{equation}
    O\left(
    \numberOfSlopes
    \numberOfLines \visibilityRatio
    \left(
    \moved{\numberOfSources}
    +
    \moved{\numberOfListeners}
    \right)
    \right)
    \, .
    \label{eq:residue dots cost}
\end{equation}
The dot product in~\eqref{eq:split residues total} is not required in practice, since the residue components can be stored and applied separately.

In addition to avoiding the time-iterative multiplications with $\FeedbackMat$, $\Ingains$, and $\Outgains$, \ac{MoD-ART} also circumvents the memory requirements of \ac{TD-ART}.
The temporal evolution of modes is implicitly described by the poles, and the \ac{EIR}'s reconstruction with~\eqref{eq:TD decomposition} only requires enough memory to store the \ac{EIR} itself.
The eigenvectors $\leftVector_\modeidx$, $\rightVector_\modeidx$ require memory equal to $2 \numberOfLines \numberOfSlopes$ (two eigenvectors of size $\numberOfLines$ for each mode).
The individual elements of $\Ingains$ and $\Outgains$ can be evaluated and applied in~\eqref{eq:split residues b}~and~\eqref{eq:split residues c} without allocating memory space for the entire matrices $\Ingains$ and $\Outgains$, avoiding an additional memory requirement of ${\numberOfLines \visibilityRatio (\numberOfSources + \numberOfListeners)}$.

Finally, as the decomposition only provides the late reverberation, the computational cost of modeling the early echoes should be taken into account.
That is to say, if accurate early reflections are desired, \ac{MoD-ART} needs to be modified by adding two extra steps in the sequence presented at the start of Section~\ref{sec:proposed}: model the early reflections using e.g. beam-tracing or ray-tracing (very small $\numberOfReflections$), and splice the early and late \acp{EIR} together.
If \ac{RTM} is chosen to do this, some operations may be saved by performing this at the same time as the ray-tracing for $\Ingains$ and $\Outgains$.

\subsection{Comparisons}
\label{sec:complexity-conclusions}

To summarize, in a scenario where all sources and listeners may move interactively, the runtime computational complexity (in terms of operations) of \ac{RTM} is given by~\eqref{eq:Tree RTM}, that of \ac{TD-ART} is the sum of~\eqref{eq:ART tracing cost} and~\eqref{eq:ART running cost}, and that of \ac{MoD-ART} is the sum of~\eqref{eq:ART tracing cost} and~\eqref{eq:residue dots cost}.
If the sound sources' positions are fixed,~\eqref{eq:ART running cost}~may be reduced to $O\left(\numberOfSamples\numberOfLines\visibilityRatio\numberOfListeners\right)$ at the expense of memory space.
These costs are compared in \figurename~\ref{fig:Complexity}.
For simplicity, we consider the same number of sources and listeners (${\numberOfSources = \numberOfListeners}$), and assume that all of them are updated at every opportunity (${\moved{\numberOfSources} = \numberOfSources}$, ${\moved{\numberOfListeners} = \numberOfListeners}$).
The plots in \figurename~\ref{fig:Complexity} include two versions of \ac{RTM}: one with a low reflection order $\numberOfReflections = 10$, which can only capture early reflections; one with a higher reflection order $\numberOfReflections = 100$, more suitable for late reverberation modeling.
These are compared with \ac{TD-ART} and \ac{MoD-ART}, as well as the special case of \ac{TD-ART} where all sources are static.
In both plots of \figurename~\ref{fig:Complexity}, we consider ${\numberOfRays = 10^5}$ rays, ${\numberOfSamples = 2000}$ \ac{EIR} samples (two seconds at \qty{1}{\kilo\hertz}), and ${\numberOfSlopes = 10}$ slopes.

In \figurename~\ref{fig:Complexity_triple}, each method's computational complexity is shown as a function of the number of sources and listeners.
For this plot, we consider the geometry shown in \figurename~\ref{fig:environment}, which is also used for the simulation results in the next section.
The scene has ${\numberOfPolys = 140}$ surface patches and ${\numberOfLines = 7982}$ volumetric paths, implying a visibility factor of ${\visibilityRatio \approx 0.4}$.
With these parameters, \ac{TD-ART} is more computationally expensive than \ac{RTM} when the number of interactive elements is low, but becomes preferable with ten or more sources and listeners.
In the special case with static sources, \ac{TD-ART} is always more advantageous than \ac{RTM}; in fact, it has lower complexity than even low-order \ac{RTM}, while capturing late reverberation much more accurately.
Running \ac{MoD-ART} requires even fewer operations, as well as allowing the movement of sources and listeners both, and having much lower memory requirements than either \ac{TD-ART} implementation.
Overall, both \ac{TD-ART} and \ac{MoD-ART} scale better with the number of sources and listeners than \ac{RTM} does.

In \figurename~\ref{fig:Complexity_visibility}, the number of sources and listeners is fixed to ${\numberOfSources = \numberOfListeners = 10}$, and complexity is shown as a function of the number of polygons.
The complexity of \ac{TD-ART} and \ac{MoD-ART} is shown for three different levels of geometrical complexity of the environment: a case with moderate visibility (${\visibilityRatio = 0.5}$: each polygon has, on average, visibility on half of all polygons), a case with low visibility (${\visibilityRatio = 0.25}$: each polygon has, on average, visibility on a quarter of all polygons), and the case of a strictly convex enclosure (${\visibilityRatio = 1}$: each polygon has visibility on all polygons).
The complexity of \ac{RTM} is unaffected by the visibility factor.
Note that having ${\visibilityRatio = 1}$ is quite unlikely in environments of interest for our application: higher levels of geometrical complexity (i.e. lower values of $\visibilityRatio$) are what lead to complex late reverberation behavior, which \ac{MoD-ART} is designed for.
Moreover, note that the upper limit of ${\numberOfPolys = 10^4}$ polygons on the axis of \figurename~\ref{fig:Complexity_visibility} is very high in the context of room acoustics modeling~\mcite{Geometry_reduction}.
Due to the logarithmic complexity scaling of ray-tracing with respect to the number of polygons, \ac{RTM} is much more efficient than either implementation of \ac{TD-ART} when the environment's polygon count is moderate-to-high.
Fully interactive \ac{TD-ART} is only preferable to \ac{RTM} in scenarios with low polygon counts and high geometrical complexity.
In contrast, \ac{MoD-ART} is in most cases more efficient than even low-order \ac{RTM}, and only surpasses the complexity of high-order \ac{RTM} in the worst-case scenarios with very high polygon counts and very low geometrical complexity.

\section{Results}
\label{sec:results}

This section demonstrates different aspects of the proposed method.
The correspondence between real, positive poles and energy decay slopes is illustrated, and the selection of modes based on their magnitude is justified, in Section~\ref{sec:results-transition}.
One possible use case for complex poles is presented in Section~\ref{sec:results-flutter}, where they are used to characterize a flutter echo.
Lastly, Section~\ref{sec:results-fractional} illustrates the effects of different \ac{EIR} sample rates and fractional delays on the decomposition.

\begin{figure}[bt]
    \centering
    \begin{adjustbox}{max width=0.45\textwidth}
    \begin{tikzpicture}[3d view = {-20}{20}]
    \newcommand \FrontOpacity {0.2};
    \newcommand \BackOpacity {0.8};
    
    \coordinate (Src) at (2.0, 2.0, 1.5);
    \coordinate (Rec1) at (2.0, 6.8, 1.5);
    \coordinate (Rec2) at (8.8, 3.5, 1.5);
    \coordinate (Rec3) at (9.3, 10.2, 1.5);
    
    \coordinate (Srcshadow) at (2.0, 2.0, 0.0);
    \coordinate (Rec1shadow) at (2.0, 6.8, 0.0);
    \coordinate (Rec2shadow) at (8.8, 3.5, 0.0);
    \coordinate (Rec3shadow) at (9.3, 10.2, 0.0);

    \coordinate (posXshift) at (0.5, 0, 0);
    \coordinate (posYshift) at (0, 0.5, 0);
    \coordinate (posZshift) at (0, 0, 0.5);
    \coordinate (negXshift) at (-0.5, 0, 0);
    \coordinate (negYshift) at (0, -0.5, 0);
    \coordinate (negZshift) at (0, 0, -0.5);
    
    \coordinate (V1) at (0.0, 0.0, 0.0);
    \coordinate (V2) at (0.0, 8.0, 0.0);
    \coordinate (V3) at (3.99, 8.0, 0.0);
    \coordinate (V4) at (4.0, 4.25, 0.0);
    \coordinate (V5) at (4.01, 4.99, 0.0);
    \coordinate (V6) at (8.5, 5.0, 0.0);
    \coordinate (V7) at (6.0, 5.01, 0.0);
    \coordinate (V8) at (6.0, 13.0, 0.0);
    \coordinate (V9) at (10.0, 13.0, 0.0);
    \coordinate (V10) at (10.0, 5.0, 0.0);
    \coordinate (V11) at (10.0, 2.0, 0.0);
    \coordinate (V12) at (4.01, 2.0, 0.0);
    \coordinate (V13) at (4.0, 2.75, 0.0);
    \coordinate (V14) at (3.99, 0.0, 0.0);
    \coordinate (V15) at (0.0, 0.0, 3.0);
    \coordinate (V16) at (0.0, 8.0, 3.0);
    \coordinate (V17) at (3.99, 8.0, 3.0);
    \coordinate (V18) at (4.0, 4.25, 3.0);
    \coordinate (V19) at (4.01, 4.99, 3.0);
    \coordinate (V20) at (8.5, 5.0, 3.0);
    \coordinate (V21) at (6.0, 5.01, 3.0);
    \coordinate (V22) at (6.0, 13.0, 3.0);
    \coordinate (V23) at (10.0, 13.0, 3.0);
    \coordinate (V24) at (10.0, 5.0, 3.0);
    \coordinate (V25) at (10.0, 2.0, 3.0);
    \coordinate (V26) at (4.01, 2.0, 3.0);
    \coordinate (V27) at (4.0, 2.75, 3.0);
    \coordinate (V28) at (3.99, 0.0, 3.0);
    
    \draw[thin, dashed, fill = myOrange!70!white, fill opacity = \BackOpacity] (V9) -- (V23) -- (V22) -- (V8) -- cycle;
    \draw[thin, dashed, fill = myOrange!50!white, fill opacity = \BackOpacity] (V10) -- (V24) -- (V23) -- (V9) -- cycle;
    \draw[thin, dashed, fill = myOrange, fill opacity = \BackOpacity] (V10) -- (V9) -- (V8) -- (V7) -- (V6) -- cycle;
    
    \draw[thin, dashed] (Rec3) -- (Rec3shadow);
    \draw (Rec3shadow) node[cross] {};
    
    \draw[thin, fill = myOrange!50!white, fill opacity = \FrontOpacity] (V8) -- (V22) -- (V21) -- (V7) -- cycle;
    \draw[thin, fill = myOrange!70!white, fill opacity = \FrontOpacity] (V7) -- (V21) -- (V20) -- (V6) -- cycle;
    \draw[thin, fill = myOrange, fill opacity = \FrontOpacity] (V22) -- (V23) -- (V24) -- (V20) -- (V21) -- cycle;
    
    \annotateLength{\qty{4}{\metre}}{V22}{V23}{posZshift};
    \annotateLength{\qty{8}{\metre}}{V23}{V24}{posZshift};
    
    \draw[thin, dashed, fill = myLightBlue!70!white, fill opacity = \BackOpacity] (V6) -- (V20) -- (V19) -- (V5) -- cycle;
    \draw[thin, dashed, fill = myLightBlue!50!white, fill opacity = \BackOpacity] (V11) -- (V25) -- (V24) -- (V10) -- cycle;
    \draw[thin, dashed, fill = myLightBlue, fill opacity = \BackOpacity] (V13) -- (V12) -- (V11) -- (V10) -- (V6) -- (V5) -- (V4) -- cycle;
    
    \draw[thin, dashed] (Rec2) -- (Rec2shadow);
    \draw (Rec2shadow) node[cross] {};
    
    \draw[thin, fill = myLightBlue!50!white, fill opacity = \FrontOpacity] (V13) -- (V27) -- (V26) -- (V12) -- cycle;
    \draw[thin, fill = myLightBlue!50!white, fill opacity = \FrontOpacity] (V5) -- (V19) -- (V18) -- (V4) -- cycle;
    \draw[thin, fill = myLightBlue!70!white, fill opacity = \FrontOpacity] (V26) -- (V25) -- (V11) -- (V12) -- cycle;
    \draw[thin, fill = myLightBlue, fill opacity = \FrontOpacity] (V24) -- (V25) -- (V26) -- (V27) -- (V18) -- (V19) -- (V20) -- cycle;
    
    \annotateLength{\qty{6}{\metre}}{V12}{V11}{negZshift};
    
    \draw[thin, dashed, fill = myGreen!70!white, fill opacity = \BackOpacity] (V3) -- (V17) -- (V16) -- (V2) -- cycle;
    \draw[thin, dashed, fill = myGreen!50!white, fill opacity = \BackOpacity] (V4) -- (V18) -- (V17) -- (V3) -- cycle;
    \draw[thin, dashed, fill = myGreen!50!white, fill opacity = \BackOpacity] (V14) -- (V28) -- (V27) -- (V13) -- cycle;
    \draw[thin, dashed, fill = myGreen, fill opacity = \BackOpacity] (V14) -- (V13) -- (V4) -- (V3) -- (V2) -- (V1) -- cycle;

    \draw[thin, dashed] (Src) -- (Srcshadow);
    \draw (Srcshadow) node[cross] {};
    \draw[thin, dashed] (Rec1) -- (Rec1shadow);
    \draw (Rec1shadow) node[cross] {};
    
    \draw[thin, fill = myGreen!50!white, fill opacity = \FrontOpacity] (V2) -- (V16) -- (V15) -- (V1) -- cycle;
    \draw[thin, fill = myGreen!70!white, fill opacity = \FrontOpacity] (V15) -- (V28) -- (V14) -- (V1) -- cycle;
    \draw[thin, fill = myGreen, fill opacity = \FrontOpacity] (V16) -- (V17) -- (V18) -- (V27) -- (V28) -- (V15) -- cycle;

    \annotateLength{\qty{3}{\metre}}{V24}{V25}{posZshift};
    \annotateLength{\qty{3}{\metre}}{V11}{V25}{posXshift};
    \annotateLength{\qty{4}{\metre}}{V1}{V14}{negZshift};
    \annotateLength{\qty{8}{\metre}}{V2}{V1}{negZshift};
    \annotateLength{\qty{2}{\metre}}{V19}{V21}{posZshift};
    \annotateLength{\qty{3}{\metre}}{V17}{V19}{posZshift};
    \annotateLength{\qty{1.5}{\metre}}{V20}{V24}{posZshift};
    \annotateLength{\qty{1.5}{\metre}}{V18}{V27}{posZshift};
    
    \fill (Src) circle (2pt);
    \fill[black, font = \footnotesize] (Src) node [right] {S};
    \fill (Rec1) circle (2pt);
    \fill[black, font = \footnotesize] (Rec1) node [right] {L1};
    
    \fill (Rec2) circle (2pt);
    \fill[black, font = \footnotesize] (Rec2) node [left] {L2};
    
    \fill (Rec3) circle (2pt);
    \fill[black, font = \footnotesize] (Rec3) node [right] {L3};
    
    \draw[->, very thick] (-0.25, 0, 0) -- (1, 0, 0) node[pos = 0.75, above]{$x$};
    \draw[->, very thick] (0, -0.25, 0) -- (0, 1, 0) node[pos = 0.9, above]{$y$};
    \draw[->, very thick] (0, 0, -0.25) -- (0, 0, 1) node[pos = 0.8, above right]{$z$};

    \coordinate (Legend) at (2, 12, 4);
    \node[draw, fill = myGreen, shape = rectangle, minimum width = 0.4cm, minimum height = 0.4cm, right = 0.0 of Legend] (Box1) {};
    \node[right = 0.2 of Box1] (Label1) {$\alpha_1 = 0.2$};
    \node[draw, fill = myLightBlue, shape = rectangle, minimum width = 0.4cm, minimum height = 0.4cm, below = 0.2cm of Box1] (Box2) {};
    \node[right = 0.2 of Box2] (Label2) {$\alpha_2 = 0.01$};
    \node[draw, fill = myOrange, shape = rectangle, minimum width = 0.4cm, minimum height = 0.4cm, below = 0.2cm of Box2] (Box3) {};
    \node[right = 0.2 of Box3] (Label3) {$\alpha_3 = 0.1$};
\end{tikzpicture}
    \end{adjustbox}
    \caption{The environment used in the presented tests. Source and listener positions are also reported in Table~\ref{tab:positions}.}
    \label{fig:environment}
\end{figure}

\begin{table}[bt]
    \caption{\itshape Source and listener positions used in the presented tests. These are also shown in \figurename~\ref{fig:environment}.}
    \centering
    \begin{tabular}{@{}ccccc@{}}
        \toprule
         & S & L1 & L2 & L3\\
        \midrule
        $x$ (\si{\meter}) & \qty{2.0}{} & \qty{2.0}{} & \qty{8.8}{} & \qty{9.3}{} \\
        $y$ (\si{\meter}) & \qty{2.0}{} & \qty{6.8}{} & \qty{3.5}{} & \qty{10.2}{} \\
        $z$ (\si{\meter}) & \qty{1.5}{} & \qty{1.5}{} & \qty{1.5}{} & \qty{1.5}{} \\
        \bottomrule
    \end{tabular}
    \label{tab:positions}
\end{table}


\begin{figure}[h!]
    \centering
    \subfloat[Listener L1]{%
        \begin{adjustbox}{width=0.48\textwidth}%
        \begin{tikzpicture}
\begin{axis}
[
width = 9cm,
height = 6cm,
scale only axis,
legend cell align = {left},
legend style = {
    at = {(0.98, 0.98)},
    anchor = north east,
    fill opacity = 0.7,
    draw opacity = 1,
    text opacity = 1,
    draw = lightgray
},
tick align = outside,
tick pos = left,
x grid style = {darkgray},
y grid style = {darkgray},
xtick style = {color = black},
ytick style = {color = black},
xlabel={Time [s]},
ylabel={Room energy response [\si{\decibel}]},
xmin = 0, xmax = 0.75,
ymin = 1e-7, ymax = 1e-1,
ymode=log,
log basis y=10,
yticklabel={\pgfmathparse{10*(\tick)}\pgfmathprintnumber[fixed]{\pgfmathresult}}
]

\addplot
[
line join=round,
very thick,
mymark={|}{1mm},
myGreen,
forget plot,
mesh/cols=2
] table [] {figures/data/Echo_triple_1_1_pole.dat};

\addplot
[
line join=round,
very thick,
mymark={x}{3mm},
myOrange,
forget plot,
mesh/cols=2
] table [] {figures/data/Echo_triple_1_2_poles.dat};

\addplot
[
line join=round,
very thick,
mymark={triangle}{5mm},
myLightBlue,
forget plot,
mesh/cols=2
] table [] {figures/data/Echo_triple_1_3_poles.dat};

\addplot [
very thick,
black,
forget plot,
mesh/cols=2
] table [] {figures/data/Echo_triple_1_TD_ART.dat};

\addplot[black, very thick]
table {%
-1 1
-2 1
};
\addlegendentry{TD-ART}
\addplot[myGreen, very thick, mark=none, smooth, legend image post style={mark=|}]
table {%
-1 1
-2 1
};
\addlegendentry{MoD-ART, $\TransitionTime=\qty{1}{\second}$ (1 pole)}
\addplot[myOrange, very thick, mark=none, smooth, legend image post style={mark=x}]
table {%
-1 1
-2 1
};
\addlegendentry{MoD-ART, $\TransitionTime=\qty{0.44}{\second}$ (2 poles)}
\addplot[myLightBlue, very thick, mark=none, smooth, legend image post style={mark=triangle}]
table {%
-1 1
-2 1
};
\addlegendentry{MoD-ART, $\TransitionTime=\qty{0.25}{\second}$ (3 poles)}

\end{axis}

\end{tikzpicture}%
        \end{adjustbox}%
        \label{fig:Echo_triple_1}%
    }
    \vfil
    \subfloat[Listener L2]{%
        \begin{adjustbox}{width=0.48\textwidth}%
        \begin{tikzpicture}
\begin{axis}
[
width = 9cm,
height = 6cm,
scale only axis,
legend cell align = {left},
legend style = {
    at = {(0.98, 0.98)},
    anchor = north east,
    fill opacity = 0.7,
    draw opacity = 1,
    text opacity = 1,
    draw = lightgray
},
tick align = outside,
tick pos = left,
x grid style = {darkgray},
y grid style = {darkgray},
xtick style = {color = black},
ytick style = {color = black},
xlabel={Time [s]},
ylabel={Room energy response [\si{\decibel}]},
xmin = 0, xmax = 0.75,
ymin = 1e-7, ymax = 1e-1,
ymode=log,
log basis y=10,
yticklabel={\pgfmathparse{10*(\tick)}\pgfmathprintnumber[fixed]{\pgfmathresult}}
]

\addplot
[
line join=round,
very thick,
mymark={|}{5mm},
myGreen,
forget plot,
mesh/cols=2
] table [] {figures/data/Echo_triple_2_1_pole.dat};

\addplot
[
line join=round,
very thick,
mymark={x}{5mm},
myOrange,
forget plot,
mesh/cols=2
] table [] {figures/data/Echo_triple_2_2_poles.dat};

\addplot
[
line join=round,
very thick,
mymark={triangle}{5mm},
myLightBlue,
forget plot,
mesh/cols=2
] table [] {figures/data/Echo_triple_2_3_poles.dat};

\addplot [
very thick,
black,
forget plot,
mesh/cols=2
] table [] {figures/data/Echo_triple_2_TD_ART.dat};

\addplot[black, very thick]
table {%
-1 1
-2 1
};
\addlegendentry{TD-ART}
\addplot[myGreen, very thick, mark=none, smooth, legend image post style={mark=|}]
table {%
-1 1
-2 1
};
\addlegendentry{MoD-ART, $\TransitionTime=\qty{1}{\second}$ (1 pole)}
\addplot[myOrange, very thick, mark=none, smooth, legend image post style={mark=x}]
table {%
-1 1
-2 1
};
\addlegendentry{MoD-ART, $\TransitionTime=\qty{0.44}{\second}$ (2 poles)}
\addplot[myLightBlue, very thick, mark=none, smooth, legend image post style={mark=triangle}]
table {%
-1 1
-2 1
};
\addlegendentry{MoD-ART, $\TransitionTime=\qty{0.25}{\second}$ (3 poles)}

\end{axis}

\end{tikzpicture}%
        \end{adjustbox}%
        \label{fig:Echo_triple_2}%
    }
    \vfil
    \subfloat[Listener L3]{%
        \begin{adjustbox}{width=0.48\textwidth}%
        \begin{tikzpicture}
\begin{axis}
[
width = 9cm,
height = 6cm,
scale only axis,
legend cell align = {left},
legend style = {
    at = {(0.98, 0.98)},
    anchor = north east,
    fill opacity = 0.7,
    draw opacity = 1,
    text opacity = 1,
    draw = lightgray
},
tick align = outside,
tick pos = left,
x grid style = {darkgray},
y grid style = {darkgray},
xtick style = {color = black},
ytick style = {color = black},
xlabel={Time [s]},
ylabel={Room energy response [\si{\decibel}]},
xmin = 0, xmax = 0.75,
ymin = 1e-7, ymax = 1e-1,
ymode=log,
log basis y=10,
yticklabel={\pgfmathparse{10*(\tick)}\pgfmathprintnumber[fixed]{\pgfmathresult}}
]

\addplot
[
line join=round,
very thick,
mymark={|}{1mm},
myGreen,
forget plot,
mesh/cols=2
] table [] {figures/data/Echo_triple_3_1_pole.dat};

\addplot
[
line join=round,
very thick,
mymark={x}{3mm},
myOrange,
forget plot,
mesh/cols=2
] table [] {figures/data/Echo_triple_3_2_poles.dat};

\addplot
[
line join=round,
very thick,
mymark={triangle}{5mm},
myLightBlue,
forget plot,
mesh/cols=2
] table [] {figures/data/Echo_triple_3_3_poles.dat};

\addplot [
very thick,
black,
forget plot,
mesh/cols=2
] table [] {figures/data/Echo_triple_3_TD_ART.dat};

\addplot[black, very thick]
table {%
-1 1
-2 1
};
\addlegendentry{TD-ART}
\addplot[myGreen, very thick, mark=none, smooth, legend image post style={mark=|}]
table {%
-1 1
-2 1
};
\addlegendentry{MoD-ART, $\TransitionTime=\qty{1}{\second}$ (1 pole)}
\addplot[myOrange, very thick, mark=none, smooth, legend image post style={mark=x}]
table {%
-1 1
-2 1
};
\addlegendentry{MoD-ART, $\TransitionTime=\qty{0.44}{\second}$ (2 poles)}
\addplot[myLightBlue, very thick, mark=none, smooth, legend image post style={mark=triangle}]
table {%
-1 1
-2 1
};
\addlegendentry{MoD-ART, $\TransitionTime=\qty{0.25}{\second}$ (3 poles)}

\end{axis}

\end{tikzpicture}%
        \end{adjustbox}%
        \label{fig:Echo_triple_3}%
    }
    \caption{\Aclp{EIR} for the listener positions in \figurename~\ref{fig:environment}, modeled with \ac{TD-ART} and \ac{MoD-ART}.
    Results are shown for three different values of $\TransitionTime$, using only real, positive poles in all cases.}
    \label{fig:Echo_triple}
\end{figure}


\subsection{Slopes and decay time threshold}
\label{sec:results-transition}

The connection between energy decay slopes and real, positive poles of \ac{ART} is validated by considering a coupled-volume environment comprising three rooms, shown in \figurename~\ref{fig:environment}.
The three rooms have absorption coefficients (from left to right) ${\alpha_1 = 0.2}$, ${\alpha_2 = 0.01}$, and ${\alpha_3 = 0.1}$, i.e. the middle room is more reverberant than the side rooms.
The plots in \figurename~\ref{fig:Echo_triple} show \ac{EIR} comparisons for three different positions of the listener, given one position of the sound source.
The source and listener positions are reported in Table~\ref{tab:positions}, and included in \figurename~\ref{fig:environment}.
The modal decomposition was performed with the Arnoldi approach, using an \ac{EIR} sample rate of ${\FSe = \qty{4}{\kilo\hertz}}$.

The results of the decomposition are shown for three different values of the decay time threshold $\TransitionTime$, selecting only real, positive poles in all cases.
Decreasing $\TransitionTime$~---~i.e. including more poles in the decomposition~---~gradually increases the accuracy of the result.
Specifically, including poles in decreasing order of magnitude prioritizes the accuracy of the late reverberation, gradually decreasing the time threshold after which the approximation is acceptable.
Note that \figurename~\ref{fig:Echo_triple_2}~and~\ref{fig:Echo_triple_3} show a gradual fade-in of the \ac{EIR}, due to the sound source and listener being located in different rooms~\mcite{Fade_in_2024}.
This is reflected by the signs of slopes in the decomposition: negative slopes detract energy from the \ac{EIR}.
Note how the signs of the slopes in \figurename~\ref{fig:Echo_triple} match those of residues in \figurename~\ref{fig:Residue_maps}, given the listeners' positions: the third-most reverberant mode is negative for L2, and the second-most is negative for L3.

\begin{figure}[tb]
    \centering
    \begin{adjustbox}{width=0.485\textwidth}%
    \begin{tikzpicture}
[
spy using outlines={rectangle, lens={xscale=2.5}, connect spies}
]
\begin{axis}
[
width = 9cm,
height = 6cm,
scale only axis,
legend cell align = {left},
legend style = {
    at = {(0.02, 0.02)},
    anchor = south west,
    fill opacity = 0.5,
    draw opacity = 1,
    text opacity = 1,
    draw = lightgray
},
tick align = outside,
tick pos = left,
x grid style = {darkgray},
y grid style = {darkgray},
xtick style = {color = black},
ytick style = {color = black},
xlabel={Time [s]},
ylabel={Room energy response [\si{\decibel}]},
xmin = 0, xmax = 0.75,
ymin = 3e-6, ymax = 3,
ymode=log,
log basis y=10,
yticklabel={\pgfmathparse{10*(\tick)}\pgfmathprintnumber[fixed]{\pgfmathresult}}
]

\addplot
[
line join=round,
very thick,
myLightBlue,
forget plot,
mesh/cols=2
] table [] {figures/data/Flutter_MoD_real.dat};

\addplot
[
line join=round,
thick,
myOrange,
forget plot,
mesh/cols=2
] table [] {figures/data/Flutter_MoD_complex.dat};


\addplot [
thick,
line join=round,
black,
forget plot,
mesh/cols=2
] table [] {figures/data/Flutter_truth.dat};

\addplot[black, very thick]
table {%
-1 1
-2 1
};
\addlegendentry{Measurement}
\addplot[myLightBlue, very thick]
table {%
-1 1
-2 1
};
\addlegendentry{MoD-ART, real poles}
\addplot[myOrange, very thick]
table {%
-1 1
-2 1
};
\addlegendentry{MoD-ART, complex poles}

\coordinate (spypoint) at (axis cs:0.5, 1e-4);
\coordinate (spyviewer) at (axis cs:0.5, 1e-1);
\spy[width=5.0cm, height=2.5cm] on (spypoint) in node [fill=white] at (spyviewer);

\end{axis}

\end{tikzpicture}%
    \end{adjustbox}%
    \caption{\Acl{EIR} of a measurement from the Arni dataset~\mcite{Arni} and two decompositions using the proposed approach, with or without complex poles. Complex poles capture the flutter echo in the environment, while the decomposition restricted to real, positive poles cannot model the oscillatory phenomenon.
    }
    \label{fig:Echo_flutter}
\end{figure}

\subsection{Complex poles}
\label{sec:results-flutter}

The significance of complex poles was validated by modeling a flutter echo.
A measurement from the Arni variable acoustics room dataset~\mcite{Arni} was considered~---~specifically, one using panel configuration 8, known to produce a flutter echo~\mcite{Flutter_Arni}.
The room was modeled as a simple shoebox, divided into ${\numberOfPolys = 32}$ surface patches, which resulted in ${\numberOfLines = 848}$ volumetric paths.
The sample rate $\FS$ of the measured \ac{RIR} is ${\FS = \qty{44.1}{\kilo\hertz}}$, while the modal decomposition was performed with an \ac{EIR} sample rate of ${\FSe = \qty{250}{\hertz}}$, using the \ac{EAI} approach and ideal fractional delays.
In \figurename~\ref{fig:Echo_flutter}, two different decomposition results are compared to the measured \ac{EIR}.
Both decompositions use decay time threshold ${\TransitionTime = \qty{0.3}{\second}}$; one uses only real, positive poles, and the other uses all (complex) poles.
As expected, the decomposition limited to real, positive poles only includes non-oscillatory modes, and therefore fails to capture the flutter echo.
On the contrary, the complex poles correctly capture the flutter.

Note that the prediction of flutter echoes is no trivial task, especially ones that carry on throughout late reverberation.
Past approaches have relied on wave-based simulations, or \ac{RTM} simulations with very high fidelity~\mcite{Flutter_barrel}.
In order to capture a long flutter echo reliably using ray-tracing, one needs to use a high reflection order and a very high number of rays.
This is because low numbers of rays and approximate scattering models lead to errors which worsen with increasing reflection orders~\mcite{RTM_errors}.
\Acl{ART} does not suffer from such errors, since its parameters inherently model the environmental characteristics which cause the flutter echo.
It follows that the same can be said of \ac{MoD-ART}.

\begin{figure}[tb]
    \centering
    \subfloat[Integer delays]{%
        \begin{adjustbox}{width=0.485\textwidth}%
        \input{figures/Poles_integer}%
        \end{adjustbox}%
        \label{fig:fs_comparison_integer}%
    }
    \vfil
    \subfloat[Fractional delays]{%
        \begin{adjustbox}{width=0.485\textwidth}%
        \input{figures/Poles_fractional}%
        \end{adjustbox}%
        \label{fig:fs_comparison_fractional}%
    }
    \caption{
    Poles obtained by \ac{MoD-ART} using different \ac{EIR} sample rates $\FSe$, with integer~\protect\subref{fig:fs_comparison_integer} or fractional~\protect\subref{fig:fs_comparison_fractional} delays.
    Axes show energy modes' $\RT$ and frequency as defined in~\eqref{eq:pole mag significance},~\eqref{eq:pole phase significance}.
    Note that the energy modes' frequencies do \emph{not} correspond to audible frequencies nor to room modes, and should instead be interpreted as fluctuations in the \ac{EIR}~---~i.e. fluctuations in the energy envelope of the \ac{RIR}.
    }
    \label{fig:fs_comparison}
\end{figure}

\subsection{Sample rate and fractional delays}
\label{sec:results-fractional}

The effects of different \ac{EIR} sample rates on poles, previously discussed in Section~\ref{sec:proposed-significance-samplerate}, are illustrated in \figurename~\ref{fig:fs_comparison}.
The plotted quantities are the $\RT$ and frequency of the poles, defined in~\eqref{eq:pole mag significance},~\eqref{eq:pole phase significance}.
Once again, it bears repeating that the poles' frequencies do \emph{not} correspond to room modes, nor to audible frequencies, as discussed in Section~\ref{sec:proposed-significance-poles}.
The plots in \figurename~\ref{fig:fs_comparison} are produced by running the decomposition for different \ac{EIR} sample rates $\FSe$, with integer or fractional delays, having fixed $\FeedbackMat$ and the propagation distances.

The first observation to be made is that, predictably, fractional delays prevent the rounding error issue entirely.
It can be seen from \figurename~\ref{fig:fs_comparison_integer} that, when integer delays are used, changing $\FSe$ leads to changes in the resulting modes~---~as previously stated, the behavior captured with higher sample rates is the physically correct one.
On the contrary, as seen in \figurename~\ref{fig:fs_comparison_fractional}, changing $\FSe$ when fractional delays are used does not cause any change in the modes.
In fact, as $\FSe$ increases, the values found with integer delays converge to the ones found with fractional delays, which further highlights how these are the correct values.

A second observation is that, for any given $\FSe$, one may only find modes up to the relative Nyquist frequency.
As a result, if it is desired to model oscillatory behavior (i.e. complex poles are to be accounted for), $\FSe$ needs to be set high enough to capture the highest oscillation frequency to be modeled~---~which, of course, increases the number of poles and the overall complexity of the approach.
In order to capture the fundamental frequency of the flutter echo in \figurename~\ref{fig:Echo_flutter}, for example, the \ac{EIR} sample rate has to be ${\FSe \ge \qty{50}{\hertz}}$.
If, however, it is only desired to model non-oscillatory modes (i.e. only real, positive poles are to be accounted for), then $\FSe$ may be set arbitrarily low~---~which is only feasible with fractional delays, due to the rounding errors caused by integer delays.





\section{Conclusions}
\label{sec:conclusions}

In this paper, we proposed a novel approach to model late reverberation in real-time interactive applications with complex acoustic environments, including situations such as coupled volumes.
The proposed method, named \acf{MoD-ART}, is highly suitable for applications with positionally dependent late reverberation, as well as multiple sound sources and/or listeners being able to move freely.
By approaching \ac{TD-ART} as a digital filter, the salient energy decay modes of the modeled environment can be found in the form of system poles and residues.
This modal decomposition provides a compact representation of the positional dependence of late reverberation, which can then be applied efficiently, requiring minimal computations at runtime.
The proposed method is more efficient than both ray-tracing and \ac{TD-ART}, as well as requiring considerably less memory space than \ac{TD-ART}.

We presented different criterions for the selection of the most salient energy decay modes, depending on the circumstance.
We showed that said selection may be scaled in favor of efficiency or accuracy; late reverberation is always prioritized, and early reverberation may also be modeled accurately (with a flexible decay time threshold) through the selection of additional modes.
We also showed that the proposed method is able to capture periodic phenomena such as flutter echoes, providing accurate modeling throughout the late reverberation.

Future work could investigate how to implement the proposed decomposition method through a set of reverberators, shared between all sources and listeners, to remove the cost of convolution.
Previous works have proposed the use of reverberators to model coupled volume environments formed by collections of rectangular rooms~\mcite{atalay2022scattering}, rectangular rooms with unevenly distributed energy absorption~\mcite{alary2024designing}, or combinations of the two~\mcite{Late_inhomogeneous}.
Basing a reverberator implementation on \ac{MoD-ART} would enable unprecedented flexibility in modeling environments with arbitrary geometry.


{\appendices

\section*{Appendix A: State-space representation}
\label{sec:appendix-SSM}

The \ac{TD-ART} state equations~\eqref{eq:ZD-ART} may be equivalently expressed in the ``expanded'' form
\begin{subequations}
\begin{align}
    \stateSpace{\Ssstate}(z) &=
    \stateSpace{\FeedbackMat}
    z^{-1} \stateSpace{\Ssstate}(z)
    +
    \stateSpace{\Ingains}
    \Ingains(z)
    \Insignal(z)
    \, ,
    \label{eq:state-space ZD-ART state}
    \\
    \Outsignal(z) &=
    \Outgains(z)
    \stateSpace{\Outgains}
    \stateSpace{\Ssstate}(z)
    +
    \Directgains(z)
    \Insignal(z)
    \, ,
    \label{eq:state-space ZD-ART output}
\end{align}
\label{eq:state-space ZD-ART}
\end{subequations}
where ${\stateSpace{\Ssstate} \in \set{C}^{\numberOfModes\!\times\!1}}$ is the state vector (the ``flattened'' content of the propagation lines), ${\stateSpace{\FeedbackMat} \in \set{R}^{\numberOfModes\!\times\!\numberOfModes}}$ is the state transition matrix (defined in the following, in~\eqref{eq:state transition matrix}), and
\begin{align}
    \stateSpace{\Ingains} &=
    \begin{bmatrix}
        \mat{0} \\ \eye
    \end{bmatrix}
    \in \set{R}^{\numberOfModes \times \numberOfLines}
    ,
    \quad \stateSpace{\Outgains} =
    \begin{bmatrix}
        \mat{0} & \eye
    \end{bmatrix}
    \in \set{R}^{\numberOfLines \times \numberOfModes}
    .
\end{align}
Note that, in~\eqref{eq:state-space ZD-ART}, the delay elements of $\Ingains(z)$, $\Outgains(z)$, and $\Directgains(z)$ are left implicit, such that $\stateSpace{\Ssstate}$ only includes the state variables related to recursive delay lines (inside the feedback loop).
The state transition matrix $\stateSpace{\FeedbackMat}$ describes the propagation of signals within the recursion loop: in other words, it describes the joint operation of both $\DelayMatArg{\feedback}(z)$ and $\FeedbackMat$ in~\eqref{eq:ZD-ART}.
In a ``true'' state-space representation, $\stateSpace{\Ssstate}$ would also include the state variables related to delay lines outside of the loop.
The ``partial'' state-space representation in~\eqref{eq:state-space ZD-ART} is chosen for simplicity, since describing the non-recursive state variables explicitly would introduce needless complexity in these equations.
The structures of $\stateSpace{\FeedbackMat}$, $\stateSpace{\Ingains}$, and $\stateSpace{\Outgains}$ are chosen such that the last $\numberOfLines$ elements of $\stateSpace{\Ssstate}$ correspond to the vector $\Ssstate$ from~\eqref{eq:ZD-ART}.
If the propagation modeled by the diagonal matrix $\DelayMatArg{\feedback}(z)$ does not include temporal spreading, the elements of $\DelayMatArg{\feedback}(z)$ are
\begin{align}
    \diag{\DelayMatArg{\feedback}(z)}
    & =
    \left[
        z^{-\delayArgs{\feedback}{1}},
        z^{-\delayArgs{\feedback}{2}},
        \dots,
        z^{-\delayArgs{\feedback}{\numberOfLines}}
    \right]
    ,
    \label{eq:delay diagonal elements}
\end{align}
where $\DelayArg{\feedback}$ are the delay line lengths.
The structure of $\stateSpace{\FeedbackMat}$ is:
\setlength\arraycolsep{2pt}
\renewcommand{\arraystretch}{0.8}
\begin{align}
    \stateSpace{\FeedbackMat} &=
    \begin{bmatrix}
        \InnerSamples_1 & \mat{0} & \mat{0} & \hdots & \mat{0} & \mat{0} & \FirstSamples_1 \\
        \mat{0} & \InnerSamples_2 & \mat{0} & \hdots & \mat{0} & \mat{0} & \FirstSamples_2 \\
        \vdots & \vdots & \vdots & \ddots & \vdots & \vdots & \vdots \\
        \mat{0} & \mat{0} & \mat{0} & \hdots & \InnerSamples_\numberOfLines & \mat{0} & \FirstSamples_\numberOfLines \\
        \LastSamples_1 & \LastSamples_2 & \LastSamples_3 & \hdots & \LastSamples_\numberOfLines & \mat{0} & \mat{0} \\
        \mat{0} & \mat{0} & \mat{0} & \hdots & \mat{0} & \FeedbackMat & \mat{0}
    \end{bmatrix}
    ,
    \label{eq:state transition matrix}
\end{align}
where
{\allowdisplaybreaks 
\begin{align}
    \InnerSamples_\lineidx &=
    \begin{bmatrix}
        0 & 1 & 0 & \hdots & 0 \\
        0 & 0 & 1 & \hdots & 0 \\
        \vdots & \vdots & \vdots & \ddots & \vdots \\
        0 & 0 & 0 & \hdots & 1 \\
        0 & 0 & 0 & \hdots & 0
    \end{bmatrix}
    \in \set{R}^{(\delayArgs{\feedback}{\lineidx} - 2) \times (\delayArgs{\feedback}{\lineidx} - 2)}
    ,
    \label{eq:inner sample transition matrix}
    \\
    \FirstSamples_\lineidx &=
    \left[
    \smashedunderbrace{
    \begin{array}{ccc}
        0 & \hdots & 0 \\
        0 & \hdots & 0 \\
        \vdots & \ddots & \vdots \\
        0 & \hdots & 0 \\
        0 & \hdots & 0
    \end{array}
    }{\lineidx - 1}
    \begin{array}{cccc}
        0 & 0 & \hdots & 0 \\
        0 & 0 & \hdots & 0 \\
        \vdots & \vdots & \ddots & \vdots \\
        0 & 0 & \hdots & 0 \\
        1 & 0 & \hdots & 0
    \end{array}
    \right]
    \in \set{R}^{(\delayArgs{\feedback}{\lineidx} - 2) \times \numberOfLines}
    ,
    \label{eq:first sample transition matrix}
    \\[20pt]
    \LastSamples_\lineidx &=
    \begin{bmatrix}
        0 & 0 & \hdots & 0 & 0 \\
        \vdots & \vdots & \ddots & \vdots & \vdots \\
        0 & 0 & \hdots & 0 & 0 \\
        1 & 0 & \hdots & 0 & 0 \\
        0 & 0 & \hdots & 0 & 0 \\
        \vdots & \vdots & \ddots & \vdots & \vdots \\
        0 & 0 & \hdots & 0 & 0
    \end{bmatrix}
    \in \set{R}^{\numberOfLines \times (\delayArgs{\feedback}{\lineidx} - 2)}
    .
    \hspace{-80pt}
    \begin{array}{r}
    \left.
    \phantom{\begin{bmatrix}
    0 \\ \vdots \\ 0
    \end{bmatrix}}
    \right\}\lineidx - 1
    \\
    \left.
    \phantom{\begin{bmatrix}
    0 \\ 0 \\ \vdots \\ 0
    \end{bmatrix}}
    \right.
    \end{array}
    \label{eq:last sample transition matrix}
\end{align}
}
Note that $\stateSpace{\FeedbackMat}$ can only be defined when the propagation delays $\DelayArg{\feedback}$ are integer, and also note that it is extremely sparse, containing a number of nonzero elements equal to ${\numberOfModes - \numberOfLines}$ plus the number of nonzero elements of $\FeedbackMat$.
Lastly, an important feature of $\stateSpace{\FeedbackMat}$ is that its eigenvalues correspond to the poles of the \ac{TD-ART} system~\mcite{Rocchesso_circulant}.

\section*{Appendix B: Derivation of $\RT(\pole_\modeidx)$}
\label{sec:appendix-T60}

Given an energy mode
${\energyResponse_\modeidx[\sampleidx] = \residue_\modeidx \pole_\modeidx^\sampleidx}$,
consider its \acs{EDC}
\begin{align}
    \operatorname{EDC}[\sampleidx]
    &=
    \frac{
        \sum_{i=\sampleidx}^{\infty}
        \abs{\residue_\modeidx \pole_\modeidx^i}
    }{
        \sum_{i=0}^{\infty}
        \abs{\residue_\modeidx \pole_\modeidx^i}
    }
    =
    \frac{
        \frac{
            \abs{\residue_\modeidx \pole_\modeidx^\sampleidx}
        }{
            1 - \abs{\pole_\modeidx}
        }
    }{
        \frac{
            \abs{\residue_\modeidx}
        }{
            1 - \abs{\pole_\modeidx}
        }
    }
    =
    \abs{\pole_\modeidx^\sampleidx}
    \, .
    \label{eq:sample edc definition}
\end{align}
Note that the values within each sum are not squared, as the mode already describes energy.
The $\RT$ is the time (measured in seconds) at which $\operatorname{EDC}[\sampleidx]$ falls below $10^{-6}$.
Since~\eqref{eq:sample edc definition} is defined w.r.t. discrete time samples $\sampleidx$ at the \ac{EIR} sample rate $\FSe$, we have
\begin{align}
    \operatorname{EDC}(\RT \FSe)
    &=
    \abs{\pole_\modeidx^{\RT \FSe}}
    =
    10^{-6}
    \\
    \RT \FSe
    &=
    \log_{\abs{\pole_\modeidx}}(10^{-6})
    =
    \frac{\ln{10^{-6}}}{\ln{\abs{\pole_\modeidx}}}
    \, .
\end{align}
Note that this derivation requires ${\abs{\pole_\modeidx} \ne 1}$: poles on the unit circle are critically stable, and have undefined $\RT$.

}

\bibliographystyle{IEEEbib}
\bibliography{Library}

\end{document}